\documentclass[12pt]{article}
\usepackage{graphicx}
\graphicspath{{Images/}}
\usepackage{color}
\usepackage{amssymb}
\usepackage{amsmath}
\usepackage{makeidx}
\usepackage{physics}
\usepackage[utf8]{inputenc}

\usepackage[a4paper,top=3cm,bottom=2cm,left=2.5cm,right=2.5cm,marginparwidth=1.75cm]{geometry}

\usepackage[colorlinks=true, allcolors=blue]{hyperref}

\newcommand{\be}{\begin{equation}}
\newcommand{\bea}{\begin{eqnarray}}
\newcommand{\ee}{\end{equation}}
\newcommand{\eea}{\end{eqnarray}}
\def \blank{\mbox{}}

\def\J{\mbox{\bf J}}

\def\x{\mbox{{\bf x}}}

\def\r{\mbox{{\bf r}}}
\def\S{\mbox{{\bf S}}}

\def\Z{\mbox{$\bf{Z}$}}
\def\W{\mbox{$\bf{W}$}}
\def\V{\mbox{$\bf{V}$}}

\def\v{\mbox{$\bf{v}$}}

\def\W{\mbox{$\bf{W}$}}
\def\u{\mbox{$\bf{u}$}}

\def\F{\mbox{$\bf{F}$}}

\def\H{\mbox{$\bf{H}$}}
\def\h{\mbox{$\bf{h}$}}

\def\J{\mbox{$\bf{J}$}}

\def\I{\mbox{$\bf{I}$}}
\def\btheta{\mbox{$\boldsymbol{\theta}$}}
\def\bpsi{\mbox{$\boldsymbol{\psi}$}}

\def\bvarphi{\mbox{$\boldsymbol{\varphi}$}}

\newcommand{\ie}{{\it i.e.}}

\begin{document}
\begin{center}

{\large {\bf Robust Forecasting using Predictive Generalized Synchronization in Reservoir Computing}}\\
\bigskip
%\end{center}
Jason A. Platt, Adrian Wong, Randall Clark\\
\bigskip
Department of Physics\\
University of California San Diego\\
9500 Gilman Drive\\
La Jolla, CA 92093, USA\\
\bigskip
Stephen G. Penny\\
Cooperative Institute for Research\\ in Environmental Sciences\\
at the University of Colorado Boulder,\\ and
NOAA Physical Sciences Laboratory\\
Boulder, CO, 80305-3328, USA \\
\bigskip
and\\
\bigskip
Henry D. I. Abarbanel~\footnote{Corresponding Author; habarbanel@gmail.com} \\
Department of Physics, and\\
Marine Physical Laboratory,\\
Scripps Institution of Oceanography,\\
University of California San Diego\\
9500 Gilman Drive\\
La Jolla, CA 92093, USA
\bigskip
\bigskip

\today

\end{center}
\newpage
%\tableofcontents
\newpage
\section*{Abstract}
Reservoir computers (RCs) are a class of recurrent neural networks (RNNs) that can be used for forecasting the future of observed time series data.  As with all RNNs, selecting the hyperparameters in the network to yield excellent forecasting presents a challenge when training on new inputs with new or previously used RCs. We analyze a method based on predictive generalized synchronization (PGS) that gives direction in designing and evaluating the architecture and hyperparameters of an RC. To determine the occurrences of PGS, we rely on the auxiliary method to provide a computationally efficient pre-training test that guides hyperparameter selection.  We provide a metric for evaluating the RC using the reproduction of the input system's {\it Lyapunov exponents} that demonstrates robustness in prediction.
\begin{center}
\noindent\rule{10cm}{0.4pt}\\
\end{center}
\vspace{0.2in}
{\bf Dynamical systems forecasting has a long history going back to 
the Kalman filter, which described the 
solution to the {\bf linear} filtering problem with discrete data.
Descendants of this linear filter were instrumental in the development of numerical weather prediction~\cite{even09} and
elsewhere~\cite{abar2021}.
A major limitation in this analysis is the need for an accurate 
model of the underlying physical processes being described. Machine learning (ML), as a ``data driven'' technique, has promise to provide forecasts without knowledge of such a physical model.  
Doing so affords many advantages including the ability to predict systems for which the physics is 
poorly understood, but for which observations are available, as well as the potential for computational speedup using dedicated physical hardware.
One of the many challenges associated with realizing the potential of 
ML in dynamical systems forecasting is the issue of selecting hyperparameters for the ML model that lead to accurate, physically meaningful, forecasts of observed quantities. In this paper we present a method for 
guiding hyperparameter selection and provide a metric for determining whether the trained network will provide physically meaningful predictions. We accomplish this in the context of reservoir computing~\cite{luk09,verstraten09,luko11,Jaeger01, Jaeger12, Jaeger04, Schrauwen07, Maass02, Wojcik04,pathak18} using the reproduction of the input system's Lyapunov exponents. }

%We provide a metric for evaluating the RC using the reproduction of the input system's {\it Lyapunov exponents} that demonstrates robustness in prediction.

%To determine the occurrences of PGS, we rely on the auxiliary method to provide a computationally efficient pre-training test that guides hyperparameter selection

\section{Introduction}
Machine Learning (ML) is a computing paradigm for data-driven prediction in which an algorithm is presented with input data in a network training phase, and then asked to predict the future behavior of new data in a generalization/forecast phase.  No detailed physical model is needed for this procedure.  When the data is in the form of a time series, the ML ``device'' is denoted a ``recurrent neural network'' (RNN)~\cite{goodfellow16}.  Different forms of RNNs used in time series prediction include GRUs \cite{GRU14} and LSTMs \cite{LSTM97}, as well as the topic of interest for this paper, reservoir computers (RC) \cite{Jaeger01}.

RNNs have feedback connectivity among the network nodes enabling self excitation as a dynamical system; this characteristic is what distinguishes it from other ML network architectures---such as multi-layer perceptrons---that assume the statistical independence of inputs~\cite{Lipton15}.  This feature identifies RNNs as an attractive choice for data driven ML forecasting~\cite{Vlachas20}.

A class of RNN architectures with demonstrated capability for forecasting dynamical systems is reservoir computing (RC)~\cite{luk09,verstraten09,luko11,
Jaeger01, Jaeger12, Jaeger04, Schrauwen07, Maass02, Wojcik04,pathak18}. In RC, a large fixed random network is selected and only the final output layer is trained---typically through linear regression.  The network is straightforward to train because the architecture and weights in the reservoir layer are fixed during training and operation, bypassing the vanishing/exploding gradients that may appear to trouble other RNN models \cite{Hochreiter01,Bengio93,Pascanu13}.  

The training input signal to the network may be generated from a known dynamical system~\cite{hunt19,ottdresden19}, or it may result from observations where the underlying dynamical rules are unknown. RC's ease of training and demonstrated prediction capabilities make them a serious contender for time series forecasting tasks~\cite{Vlachas20}.
%fig 1
\begin{figure}[!htpb]
    \centering
    \includegraphics[width = 1.02\textwidth,height=0.65\textwidth]{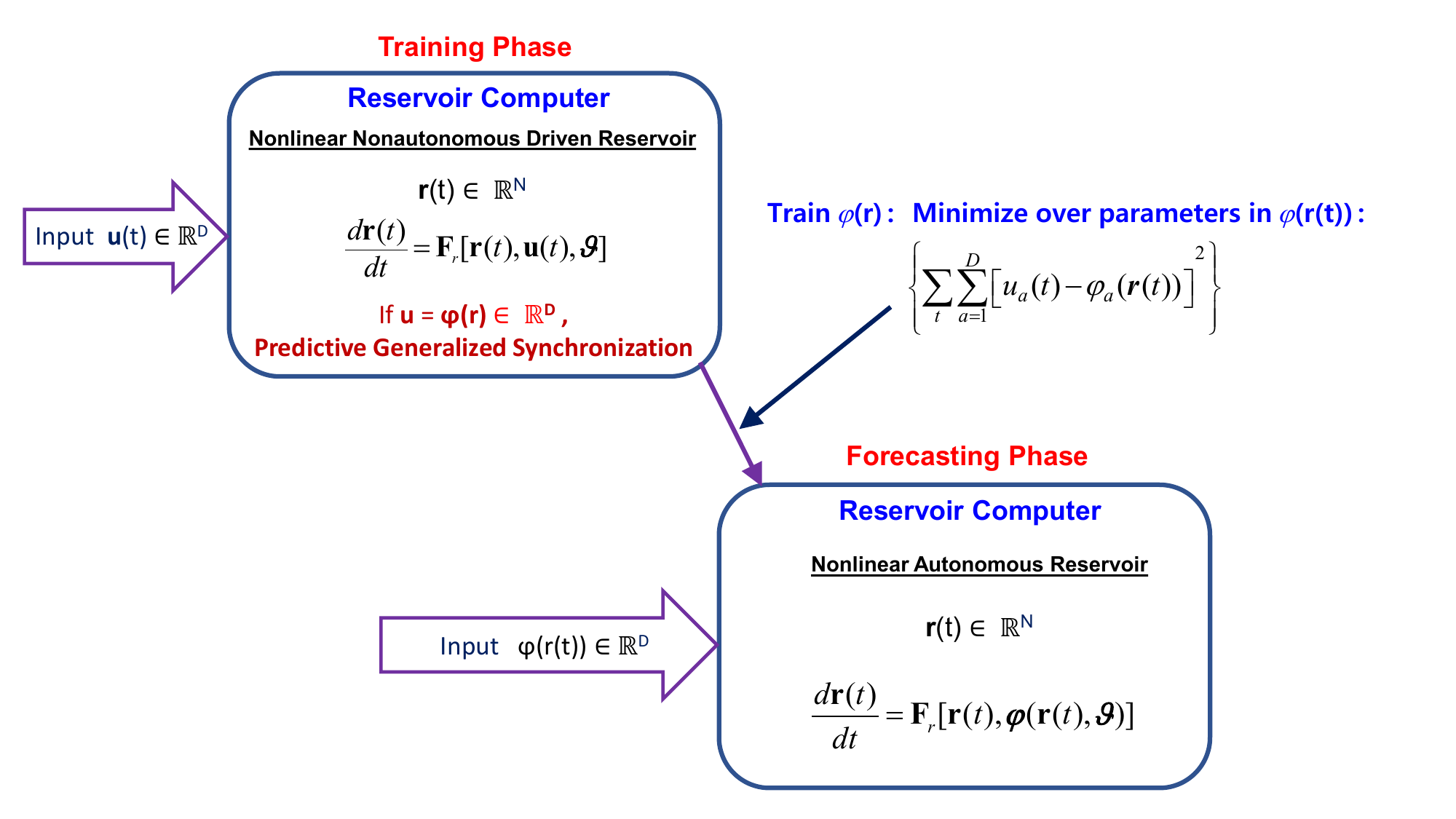}
    \caption{Flow of operations for implementing a Reservoir Computation (RC) strategy to perform forecasting/prediction of an input  $\u(t) \in \mathbb{R}^D$ presented to a RC with dynamical degrees-of-freedom  $\r(t) \in \mathbb{R}^N$. The RC dynamics are given as $\dot{\r}(t) = \F_r(\r(t), \u(t),\btheta)$; $\btheta$ are fixed parameters in the RC. When the input and the reservoir exhibit {\it predictive generalized synchronization} (PGS), $u_a = \varphi_a(\r); a = 1,2,...D$; training consists of estimating any parameters in a representation of $\bvarphi(\r)$. After the regions of PGS are established for a given $\u(t)$ and a selected $\F_r(\r,\u,\btheta)$, one may wish to change the values of $\btheta$ within the PGS region to optimize the predictive performance of the RC.}
    \label{fig: res_pic}
\end{figure}

The ability to develop a data-driven model using a method such as RC is attractive for a number of practical reasons. RC allows us to construct predictive models of unknown or poorly understood observed dynamics.   Should the input signal $\u(t)$ arise from measurements of high dimensional geophysical or laboratory flows~\cite{Sharan19,Matheou18}, for example, the striking speedup in computing with a reservoir network realized in hardware~\cite{Tanaka19,canaday18} may permit the exploration of detailed statistical questions about the observations that might be difficult or impossible otherwise. RC has the potential to provide significant computational cost savings in prediction applications, since the RC dynamics typically comprise a network with computationally simple active dynamics at its nodes.

\subsection{Discerning how RC Works}\label{discern}
A reservoir is a dynamical system with coordinates $\r =\{r_1,r_2,...r_N\}$ which is driven by a signal $\u = \{u_1,u_2,...u_D\}$ where $D \ne N$; usually $D \ll N$. The dynamical equation of the reservoir is represented by a recurrent network as (~\cite{goodfellow16}, Section 10.4)
\be
\frac{dr_{\alpha}(t)}{dt} = F_{\alpha}(\r(t),\u(t),\btheta);\;\alpha = 1,2,...N.
\label{rnn}
\ee
$\btheta$ are parameters in the recurrent network (reservoir)
The goal of reservoir computing is to have the reservoir `learn' the properties of the training signal $\u(t)$, and, having been trained, act as a useful prediction machine for the future of the training signal.

The success of RNNs and their increased adoption in research applications has perhaps out paced the understanding of how these data driven processes are successful. It is not known how best to design a network for a particular problem, nor how much or what kind of data is most useful for training. General guidelines are established and widely discussed~\cite{vers07,Schrauwen07,luko11,Jaeger12}. These tend to be justified with empirical rather than theoretical considerations. In probing this question, the very interesting idea arose~\cite{hunt19,lymburn19} that the explanation might be a form of synchronization known as `generalized synchronization' (GS)~\cite{ sushchik95,ars96,kocarev96}.

Unidirectional driven systems such as these have long been studied, especially in the analysis of nonlinear dynamical systems~\cite{abar96,kantz04}. There exists a kind of synchronization between systems with different numbers of state variables, such as the D-dimensional source of the $\u(t)$ and the N-dimensional reservoir $\r(t)$, and this is called generalized synchronization (GS)~\cite{sushchik95,ars96}; this is in contrast with the synchronization introduced by Pecora and Carroll~\cite{pc90}, which analyzes how two identical systems may synchronize.  GS is a statement that there is a relation $r_{\alpha}(t) = \bpsi_{\alpha}(\u(t))$ meaning the $\r(t)$ dynamics tracks the driving dynamics $\u(t)$ even though they can be quite different dynamical systems.

GS was established by experiments as early as 1998~\cite{tang98} and has a role in areas as diverse as synchronization of neurons in a brain circuit to cryptographic code breaking\cite{Alvarez06}. In the synchronization of dynamical systems in the presence of noise, the two systems being synchronized are never precisely the same so GS is essential in explaining how synchronization works in practice.

In the framework of reservoir computing---to achieve the goal of the reservoir learning the dynamical information in the signal---one uses the idea that the signal $\u = \W_{out} \r$, here $\W_{out}$ is a $D \times N$ matrix to be estimated, turns the driven system Eq. (\ref{rnn}) into an autonomous system. The reservoir may be analyzed as an initial value problem while yielding a forecasting model for the learning $\u(t)$. In the language of GS this suggests that the {\em inverse} of the GS relationship
$r_{\alpha}(t) = \bpsi_{\alpha}(\u(t))$, namely $u_a(t) = \varphi_a(\r(t));\;a=1,2,...,D$ would generalize the linear $\u \to \r$ feedback relation $\u = \W_{out} \r$.

The existence of the inverse GS relationship is associated with the requirements of the inverse function theorem and requires the functions $\bpsi(\u)$ and $\bvarphi(\r)$ be differentiable \cite{Hunt97}. This is a local requirement for the inverse of $\bpsi(\u)$ to exist.

Dynamical systems often have many basins of attraction in $\r$ space, and the basin one is in depends on the choice of $\r(0)$ in the initial value solution of the reservoir equation. The boundaries between basins of attraction are known to be fractal in many examples~\cite{mc85}, and maintaining differentiability across such boundaries would appear to be unlikely. The idea that $\bpsi(\u)$ is invertible, expressed as invertible GS \cite{Lu20}, cannot be a global property of all reservoirs. The inverse in one basin of attraction may be quite different in another.

% In the analysis of RCs we need a modification of $\r(t) = \bpsi(\u(t))$ to the relation $\u(t) = \bvarphi(\r(t))$. To distinguish the use here of $\u = \bvarphi(\r)$ from GS we refer to $\u = \bvarphi(\r)$ as {\it predictive generalized synchronization} (PGS).  

We are interested in investigating the role of GS in the formulation of the attractive idea of RC, since a relation $\u=\bvarphi(\r)$ would turn the reservoir into an autonomous system and achieve the training needed to make the reservoir an excellent forecasting machine.

GS can be identified for a given driving or teaching signal $\u(t)$ in a most straightforward manner using the auxiliary system approach~\cite{ars96} which for choices of parameters of the reservoir selection identifies when
$r_{\alpha}(t) = \bpsi_{\alpha}(\u(t))$ holds. The condition is succinctly stated as requiring that the Lyapunov exponents of the reservoir, conditioned on the teaching signal, be negative~\cite{pc90}. This feature of GS appears equivalent to the fading memory feature of useful reservoirs.

In reservoir computing a GS relation $r_{\alpha}(t) = \bpsi_{\alpha}(\u(t))$ is not immediately useful for moving the driven reservoir into an autonomous dynamical system for $\r(t)$. We introduce the term predictive generalized synchronization (PGS) to remind us that invertibility of the GS relation is not globally true in general, and to give us a reservoir parameter region where it could hold. In such a region we may assume that $\u = \varphi(\r)$ happens and proceed with the framework of reservoir computing with confidence.

\subsection{Goals of This Paper}

The goal of this paper is to provide a useful path for finding reservoir parameters where good generalization/forecasting is possible.

The training of an RC, as indicated in Fig.(\ref{fig: res_pic}), consists of minimizing
\be
\sum_t \sum_{a=1}^D (u_a(t) - \varphi_a(\r(t))^2,
\label{rctrain}
\ee 
with respect to parameters that enter in the representation of the vector valued function $\bvarphi(\r)$ of the reservoir variables $\r(t)$. This training of the PGS function $\bvarphi(\r)$ to match the data stream $\u(t)$ is the essential step of transferring the information content in the data $\u(t)$ to the reservoir dynamics. 

% There may separately be the matter of selecting some of the global parameters, called $\btheta$ in the discussion, in the RNN dynamics of the RC, and we will take this issue up in Section (\ref{sec: testing_pgs}). 

We pursue the interesting suggestion given in~\cite{hunt19,lymburn19} that GS is instrumental in RC, using, however, PGS rather than GS, to move from a somewhat {\it ad hoc} training approach to a systematic strategy.  In training, we ensure that the input $\u(t)$ and the reservoir degrees-of freedom $\r(t)$ satisfy PGS, $\u(t) = \bvarphi(\r(t))$, and point out that it is parameters in the representation of the function $\bvarphi(\r)$ that we need to estimate.

We demonstrate a computationally efficient way to choose regions of RC hyperparameters where PGS occurs as well as regions where PGS does {\bf not} occur. This is then employed in guiding hyperparameter choices for skillful forecasting of the input training data $\u(t)$, given an accurate enough approximation to $\varphi(\r)$. These hyperparameters may include some properties of the $N \times N$ adjacency matrix $A_{\alpha,\beta}$ such as the spectral radius and the density of connections among the $\r \in {\mathbb R}^N$ active units $\rho_{A}$. These quantities are collected together in Table 1 as they appear in an RC with $\tanh$ dynamics at its nodes.

A related issue is that the traditional method of evaluating the effectiveness of an RNN, with training and testing data sets, is awkward when performing dynamical systems forecasting. This method gives no indication of the stability of the forecast.  Another approach to evaluation, showing a prediction of a single time series, also gives no indication of the stability of the predictions over the entire range of inputs.  In this paper we attempt to rectify these deficiencies by using dynamical properties of the reservoir to design and evaluate a trained network.

\subsection{Presentation Strategy in This Paper}~\label{strategy}

\begin{enumerate}
\item We introduce a computationally efficient numerical test, based on PGS, and using the `auxiliary method', to guide hyperparameter selections in RCs resulting in very good forecasting.
\item We portray the ideas for the use of PGS with some simple illustrative models~\cite{lor63,lor96,lorman98}, then discuss an important geophysical model, the Shallow Water Equations (SWE)~\cite{sadourny75,pedlosky1986,vallis17}, and finish with a discussion of a biophysical model of neuron dynamics~\cite{jwu,willshaw}. The last item comprises data from a driven dynamical system (the neuron), and these data depend on an injection of current to stimulate the neuron into interesting oscillations. The RC must obtain information about the driving force as it is trained.
\item We explore a metric for a ``well trained'' RC network using the reproduction of the input system's Lyapunov exponent spectrum. We introduce a criterion for excellent forecasting connected to the conditional Lyapunov exponents~\cite{pc90} of the reservoir, and
\item We briefly address, and speculate about, the long-term value of RC for problems encountered in physical systems.
\end{enumerate}

\section{PGS in Reservoir Computing}
RCs are often applied to forecasting problems where a familiar task is to learn from an input sequence $\u(t) \in {\mathbb R}^D$ generated from observed data produced by a likely unknown autonomous dynamical system
\be 
\frac{d\u(t)}{dt} = \F_u(\u(t)),
\label{autondatasource}
\ee
and then forecast the future of $\u(t)$.

In this paper we will utilize data from three autonomous dynamical systems:
\begin{enumerate}
\item The Lorenz 1963 three dimensional model~\cite{lor63}. This a familiar test bed for ML forecasting of chaotic dynamics.
\item The Lorenz 1996~\cite{lor96} D-dimensional model. Often used in geophysics as a platform for examining ideas in a context in which the dimension D of the produced data can be easily made large.
\item The shallow water equations (SWE). These describe the flow of a thin layer of fluid (say the ocean/atmosphere system which has a depth approximately 10-15 km) on a sphere (the earth with a radius of about 6400 km). The SWE are a set of three partial differential equations in three state variables $\{u(x,y,t),v(x,y,t),h(x,y,t)\}$  on a mid-latitude tangent plane to the earth.
\end{enumerate} 

When one has only partial information on $\u(t)$, the use of time delay embedding as described in the Appendix (\ref{tdelay}) allows the construction of a proxy space~\cite{abar96,kantz04} in which to proceed.

The reservoir is a non-autonomous dynamical system \be \frac{d\r(t)}{dt} = \F_r(\r(t),\u(t),\btheta) \ee with a vector field $\F_r(\r,\u,\btheta)$ composed of $N$ nodes in a network at which we locate nonlinear dynamical models. In ML these are often called `activation functions',~\cite{Wojcik04, Brunner16, Chrisantha03,Haynes14,Larger17,Paquot11}.  The nodes in the network are connected through an $N \times N$ adjacency matrix $A_{\alpha\,\beta}; \alpha,\beta = 1,2,...N$, chosen to have a connection density $\rho_{A}$. In addition to $\rho_A$, the largest eigenvalue of $A_{\alpha\,\beta}$, called the spectral radius (SR), is adjustable and may often be determinative of PGS.  The adjacency matrix is chosen at random with only a few global characteristics set, none of the internal weights are adjusted or trained after $A$ is generated.  Each choice for the activation functions has other parameters $\btheta$, including ones that establish the time scale for the operation of the RC and for the strength of the coupling of the data $\u(t)$ into the RC. (For a $\tanh$ reservoir, please see Table (1)).

The input maps the signal $\u(t)$ from $D$ dimensions into the $N$ dimensional reservoir space of $\r(t)$.  The output layer is a function such that $\varphi_a(\r)= u_a(t),$ chosen during the {\bf training phase} during which we estimate any parameters in $\varphi(\r)$. This is the only part of the reservoir computer that is trained.   It is common practice, followed in this paper, to choose $\bvarphi(\r)$ as a linear (or at most quadratic) function of $\r$, but this is by no means the only, or best, representation~\cite{silverman86,buhmann09,casdagli89,billings13,broom88,guill98, Scott05}. Using the selected approximation to $\bvarphi(\r)$ we develop an autonomous predictive reservoir based on learning $\u = \bvarphi(\r)$.

% The structure of an RC in training and forecasting is shown in Fig.(\ref{fig: res_pic}). Through $\bvarphi(\r)$
%  $\dot{\r}(t) = \F_r(\r,\u,\btheta)$ $\r(t) \in {\mathbb R}^N$ represents the information in the input time series  $\{\u(t_m)\}$ for $t_m = t_0 + m \Delta t;\; m = 0,1,2,...M$ but now in $N > D$ dimensional space.

\subsection{Synchronization and Training}
Generalized Synchronization refers to the synchronization of two {\bf nonidentical} dynamical systems which, by definition, cannot exhibit identical oscillations ~\cite{sushchik95,ars96,kocarev96,Pyragas98}.  For an input system $\u(t)$ and response system $\r(t)$, if there is GS, then there is a function $\bpsi$ such that $\r = \bpsi(\u)$; note that this relation is only true asymptotically---\ie, after a certain finite amount of time has passed for transients to die out. The dynamics of the response system are therefore entirely predictable from the history of the input; in RC, for a contracting system, this property is related to the ``echo state property'' \cite{Jaeger01} and ``fading memory'' \cite{boyd85}. One rarely has an explicit form for $\bpsi$, but if there is GS, then that suggests it exists~\cite{abar96,hunt19}.  To our knowledge, the details of the mathematical properties of $\bpsi(\r)$ are not known. Important information about $\bpsi$ is found in~\cite{stark97,stark99,josic00}.

When $\u(t)$ and $\r(t)$ are synchronized, the combined system in $\mathbb{R}^{N+D}$ will lie on an invariant {\it synchronization manifold} $\mathcal{M}$~\cite{Pecora97}.  $\mathcal{M}$ must be locally attracting, that is the Lyapunov exponents transverse to the manifold, called conditional Lyapunov exponents, are negative. The stability of the motion on such a manifold has been the subject of numerous inquiries \cite{Pecora97, Pecora00} but can overall be summarized by the statement that $\mathcal{M}$ must be normally hyperbolic---\ie, the contraction normal to $\mathcal{M}$ is larger than the contraction tangential to $\mathcal{M}$ \cite{Kocarev00}.

While GS is defined to be $\r = \psi(\u)$, we have defined PGS to be $\u = \varphi(\r)$ without assuming $\varphi = \psi^{-1}$ in a global sense.  PGS assumes that the system is contained in a local region where $\psi$ is smooth, invertible and differentiable---see \cite{Hunt97} for a discussion of conditions where this might hold.  If the basins of attraction to the synchronization manifold are fractal then different initial conditions of the reservoir may end up on different attractors.

PGS gives us some advantage in the analysis of RC networks; it assures us that the dynamical properties that the dynamical properties of the stimulus $\u(t)$ and the reservoir $\r(t)$ are now essentially the same. They share global Lyapunov exponents~\cite{ose68}, attractor dimensions, and other quantities classifying nonlinear systems ~\cite{sushchik95}.  The principal power of PGS in RC is that we may replace the initial non-autonomous reservoir dynamical system with an autonomous system operating on the synchronization manifold.
 
The function $\bvarphi(\r)$ is approximated in some manner,  through training, Eq. (\ref{rctrain}), and then this is substituted for $\u$ in the reservoir dynamics. In previous work on this~\cite{hunt19,ottdresden19} the authors approximated $\bvarphi(\r)$ as a polynomial expansion in the components $\r_{\alpha}$ and used a Tikhonov-Miller~\cite{miller70,tikhonov77, Press-Flannery-2007-NumRecipes} regularization method, also known as ridge regression~\cite{khalaf05}, to find the coefficients of the powers of $\r_{\alpha}$. In this paper we follow their example. However, we note that there are a large number of approximation methods for representing functions of many variables~\cite{silverman86,p-rbfmir-87,buhmann09,casdagli89,billings13,broom88,Scott05,guill98}. Some may provide a useful general representations of $\bvarphi(\r)$ whose value could exceed that of a Taylor series expansion.

The reservoir dynamics acts at the nodes of the {\it non-autonomous} network equations for $\r(t)$ given by
\be
\frac{ d\r(t)}{dt} = \F_r[\r(t), \u(t),\btheta],
\label{RNN}
\ee
or the equivalent statement of the dynamics in discrete time.

If PGS takes place, we may replace Eq. (\ref{RNN}) by the {\it autonomous} dynamical system
\be
\frac{d\r(t)}{dt} = \F_r[\r(t),\bvarphi(\r(t)),\btheta].
\ee
We use this for forecasting. What information we have about the manner in which driving forces influence the forecasting capability of the RC is now encoded in $\bvarphi(\r)$ through the PGS relation $\u = \bvarphi(\r)$.

\subsection{The Auxiliary Method for PGS}
There are a variety of approaches for determining whether one's selection of $\r(t)$ and $\u(t)$ exhibit PGS. Perhaps the most direct approach is to work with {\bf two identical} reservoirs~\cite{abar96,ars96} driven by the same $\u(t)$, 
\be
\frac{d\r_A(t)}{dt} = \F_r(\r_A(t), \u(t),\btheta) \; \mbox{and} \; \frac{d\r_B(t)}{dt} = \F_r(\r_B(t), \u(t),\btheta).
\ee
Solve these two equations, keeping $\u(t)$ fixed and with $\r_A(0) \ne \r_B(0)$; then examine the distance between $\r_A(t)$ and $\r_B(t)$ as t becomes large. If $\r_A(t) \to \r_B(t)$ we have PGS.  This `auxiliary method' was first used in an experimental verification of GS by~\cite{tang98}.

The auxiliary method is simply a restatement of the fact that GS is contingent upon the conditonal Lyapunov exponents (CLE) being negative.  The CLEs give the rate of error growth of the response system dynamics on $\mathcal{M}$.  Given a small perturbation to the RC trajectories---$\hat \r(t) = \r(t) + \delta \r(t)$---the error growth rate after linearization is
\be
\dv{t} \delta \r(t) = \F_r(\r(t), \u(t),\btheta) - \F_r(\hat \r(t), \u(t),\btheta) \Rightarrow \dv{t} \delta \r(t) = D_rF(\r(t), \u(t),\btheta) \cdot \delta \r(t),
\ee
with $D_rF$ the jacobian with respect to $\r$.  The CLEs can be calculated by solving this variational equation; for a reservoir dimension of several thousand this calculation can be computationally intensive.

\begin{figure}[!htpb]
    \centering
    \includegraphics[width=0.495\textwidth]{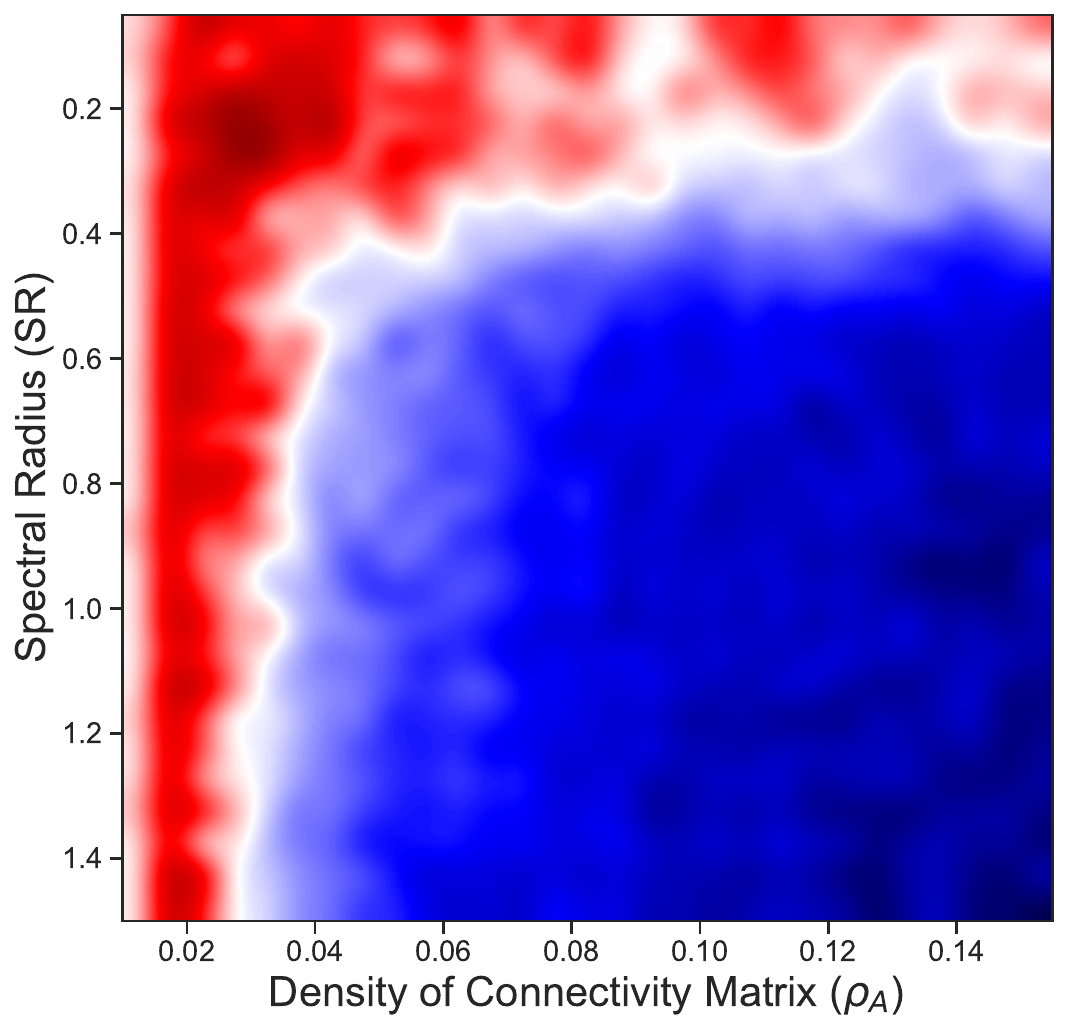}
    \includegraphics[width=0.495\textwidth]{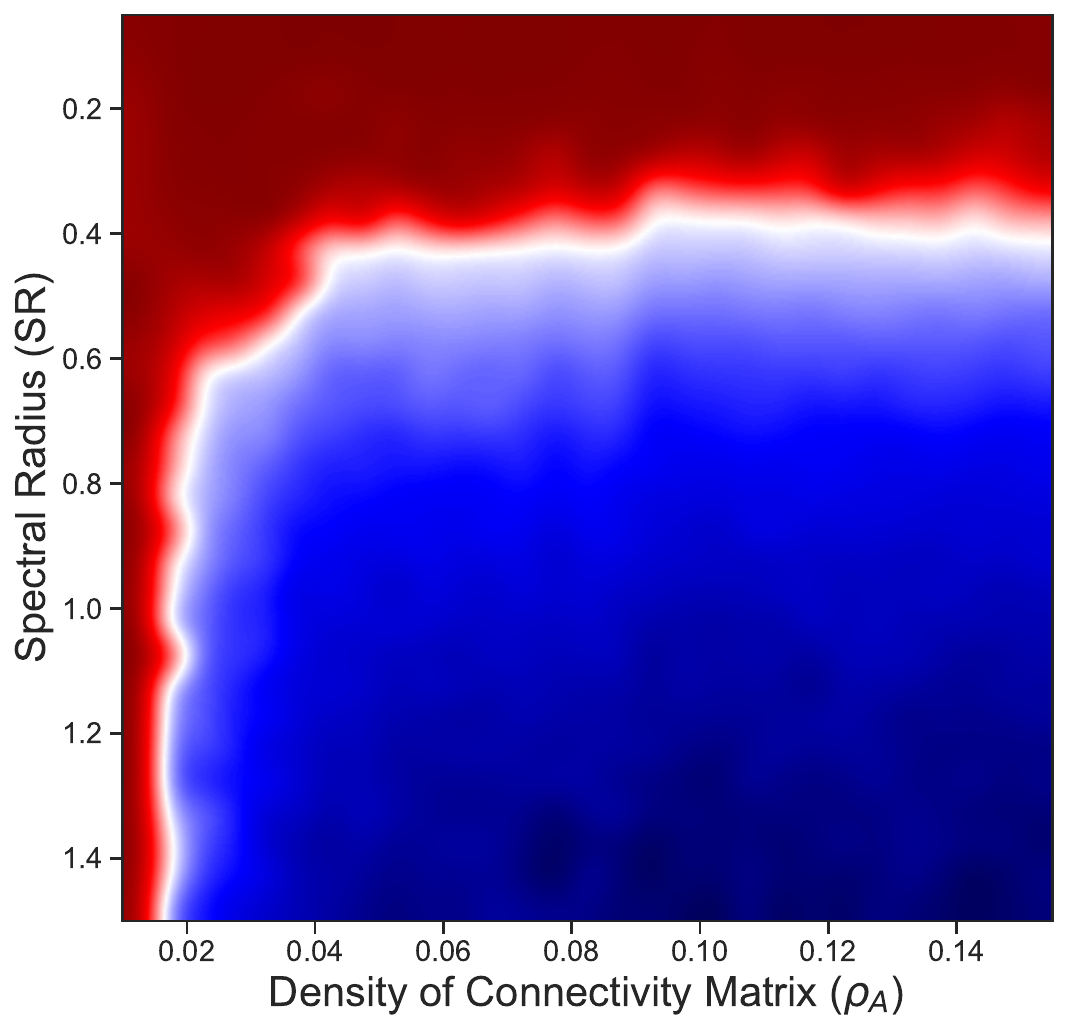}
    \caption{We display two ways of computing regions of PGS for a reservoir  with Fitzhugh-Nagumo neurons~\cite{fitzhugh1961impulses,Nagumo1962} at the nodes (N = 500).  Both methods give approximately the same result.\\ {\bf Left Panel} The largest conditional Lyapunov exponent (CLE) calculated for Lorenz63 input and a Fitzhugh-Nagumo RC as we vary the hyperparameters SR and $\rho_A$.  {\bf Blue} shows regions with {\it positive} CLEs, so no PGS. {\bf Red} shows  regions of {\it negative} CLE meaning PGS exists in this region.\\  {\bf Right Panel} The error between the response system and the {\it auxiliary} response system as $t \to \infty$: $\|\r_A(t) - \r_B(t)\|$. If this remains large, there is no PGS. If it goes to zero, there is PGS. Choices for hyperparameters in the {\bf Blue} regions indicate the absence of PGS, while choices in the {\bf Red} regions show PGS.\\ There is a slight discrepancy between the two calculations.  Part of this is certainly caused by numerical errors in estimating the CLEs of a high dimensional dynamical system using a finite trajectory. It is also possible that the auxiliary method initialized at different points could be falling into different basins of attraction resulting in a failed test even though the CLEs may be negative.}
    \label{fig: AUXandCLE_FHN_pnzSR}
\end{figure}

The advantage of the auxiliary method test is that it fast and efficient, unlike the challenges associated with evaluating the CLE directly~\cite{ose68,eckmann85,abar96}. While the results of the two methods may be essentially the same, the required computations might be quite unequal. We show the outcomes of both approaches in Fig.(\ref{fig: AUXandCLE_FHN_pnzSR}).  The CLEs~\cite{pc90} of the reservoir systems being negative mean that the two reservoir states should converge exponentially towards each other.

The auxiliary method is not a direct test for PGS.  Rather it is a test that the GS function $\psi$ exists, is smooth and continuous \cite{ars96}.  While we do not claim that these properties always guarantee that the auxiliary method implies PGS, it is highly suggestive that it does.  In addition, there are some limitations in general on the accuracy of the auxiliary method in certain circumstances.  The  method is only applicable when the driven system does not trend to a fixed point or a limit cycle for arbitrary u, but is instead ``driven'' by the input.  Moreover, one may construct systems for which a function $\varphi$ exists but for which the auxiliary method will fail.  Therefore this test will not always guarantee PGS.

To restate the claim: if the auxiliary method passes then GS exists and the synchronization manifold in all likelihood is locally smooth and differentiable implying the existence of PGS.  If the auxiliary test fails then in general GS is not occurring except in certain circumstances such as the presence of multiple basins of attraction.  Therefore PGS is almost certainly not occurring. In factorable systems (such as an RC with block diagonal entries in the adjacency matrix) all independent components would have to be tested separately.  In our experience, however, the auxiliary method is a useful practical tool for standard formulations of RC.

\subsection{Testing for PGS \label{sec: testing_pgs}}

PGS provides us with a test of whether a particular reservoir, with a specific choice of architecture, nodal dynamics and hyperparameters, has the capability to learn the dynamics presented by the data.  As described in the previous section, {\bf without training} one can simply evolve $\dot{\r}(t) = \F_r(\r(t), \u(t),\btheta)$ with the input $\u(t)$ present for two different initial conditions, and then test to see if PGS occurs.  If PGS {\bf does not occur} between the reservoir and the data, then our forecasting results will be unwelcome, and a more suitable choice of hyperparameters should be sought. 

%fig 2
\begin{figure}[!htpb]
\centering
\includegraphics[width=0.65\textwidth]{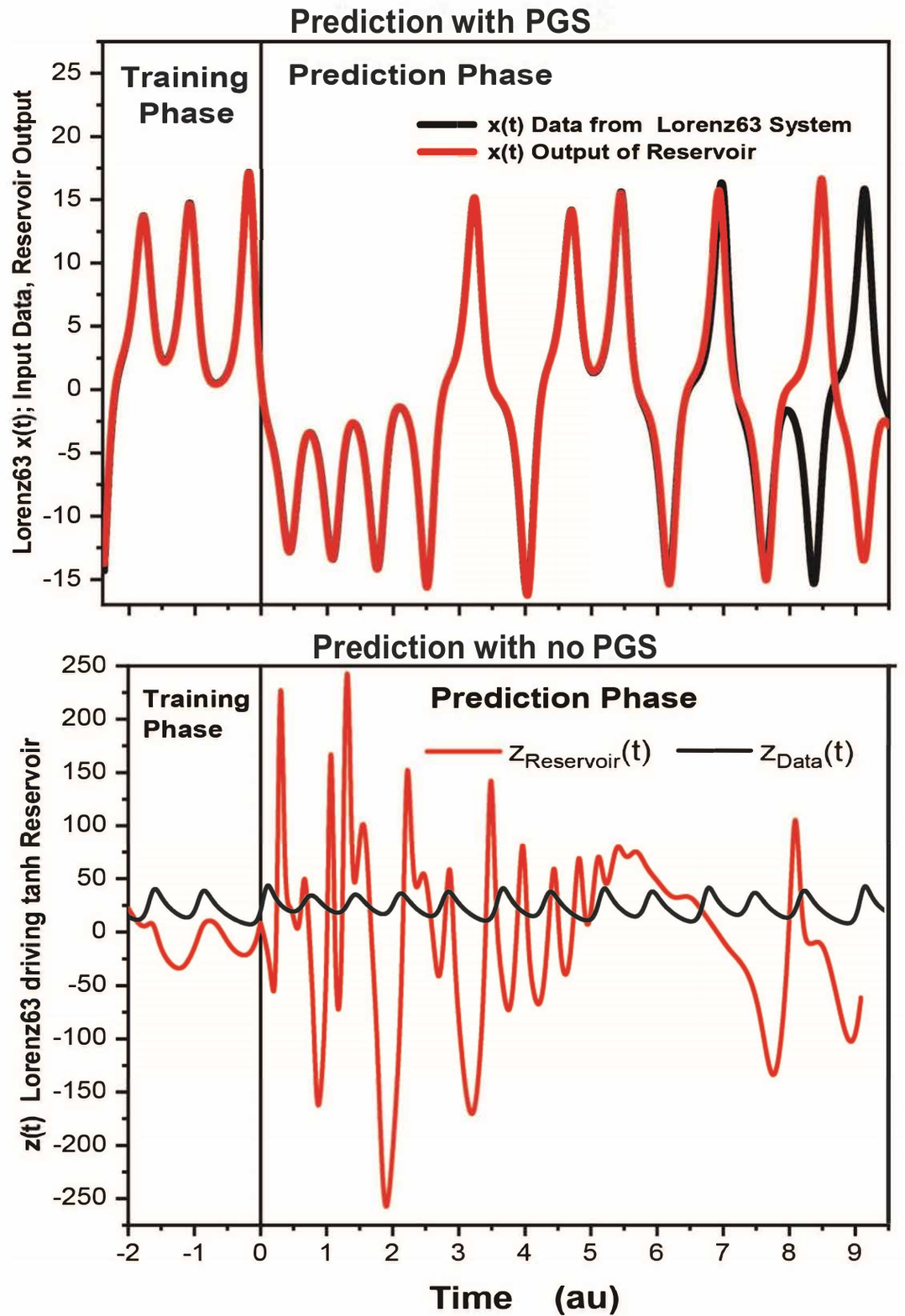}
\caption{{\bf Top} Prediction by an $N = 2000$ $\tanh$ reservoir output (red) receiving input from Lorenz63 dynamics (black) \cite{lor63}. In $A_{\alpha\,\beta}:SR = 0.9$ and $\rho_{A}$ = 0.02.  This is in the PGS region.
    The black vertical line is the end of the ``training period''. {\bf Bottom} When one selects the hyperparameters {\bf outside} the region of PGS, for example using $N = 2000$, $SR = 1.6$ and $\rho_{A} = 0.02$ for the $\tanh$ reservoir, the function $\bvarphi(\r)$ does not exist. We may expect the reservoir to operate poorly in producing a replica of the input $\u(t)$, as we can see it does.}
    \label{xlor63tores}
\end{figure}

In Fig.(\ref{xlor63tores}) we display the predictive consequences of a set of $\btheta$ that admits PGS, ${\bf Top Panel}$, and after that the predictive consequences of a set of $\btheta$ that does not admit PGS, ${\bf Bottom Panel}$. This Figure comes from a $\tanh$ reservoir driven by data coming from solving the Lorenz63~\cite{lor63} equations.

Establishing regions of $\btheta$ for PGS can greatly reduce the number of RCs with different hyperparameters that must be trained and tested in a grid search over $\btheta$ or other identification technique. 
%As may be seen in Fig.(\ref{fig: AUXandCLE_FHN_pnzSR}), the boundary between PGS and nonPGS regions is, in practice, quite sharp. So the search for PGS and nonPGS regions in $\btheta$ space may be coarse grained to begin and subsequently refined if desired.
%fig 3

\subsection{Advantages}

The advantage of searching first for PGS comes from the fact that the auxiliary method test is fast and efficient. In practice this property often means that one can look at a much smaller segment of time in the data series than is required for accurate results for training.  In addition, the linear regression step does not need to be completed, so searching for PGS is computationally much more efficient than training a reservoir and then evaluating it by predicting at multiple points. 

\begin{figure}
    \centering
    \includegraphics[width=0.6\textwidth]{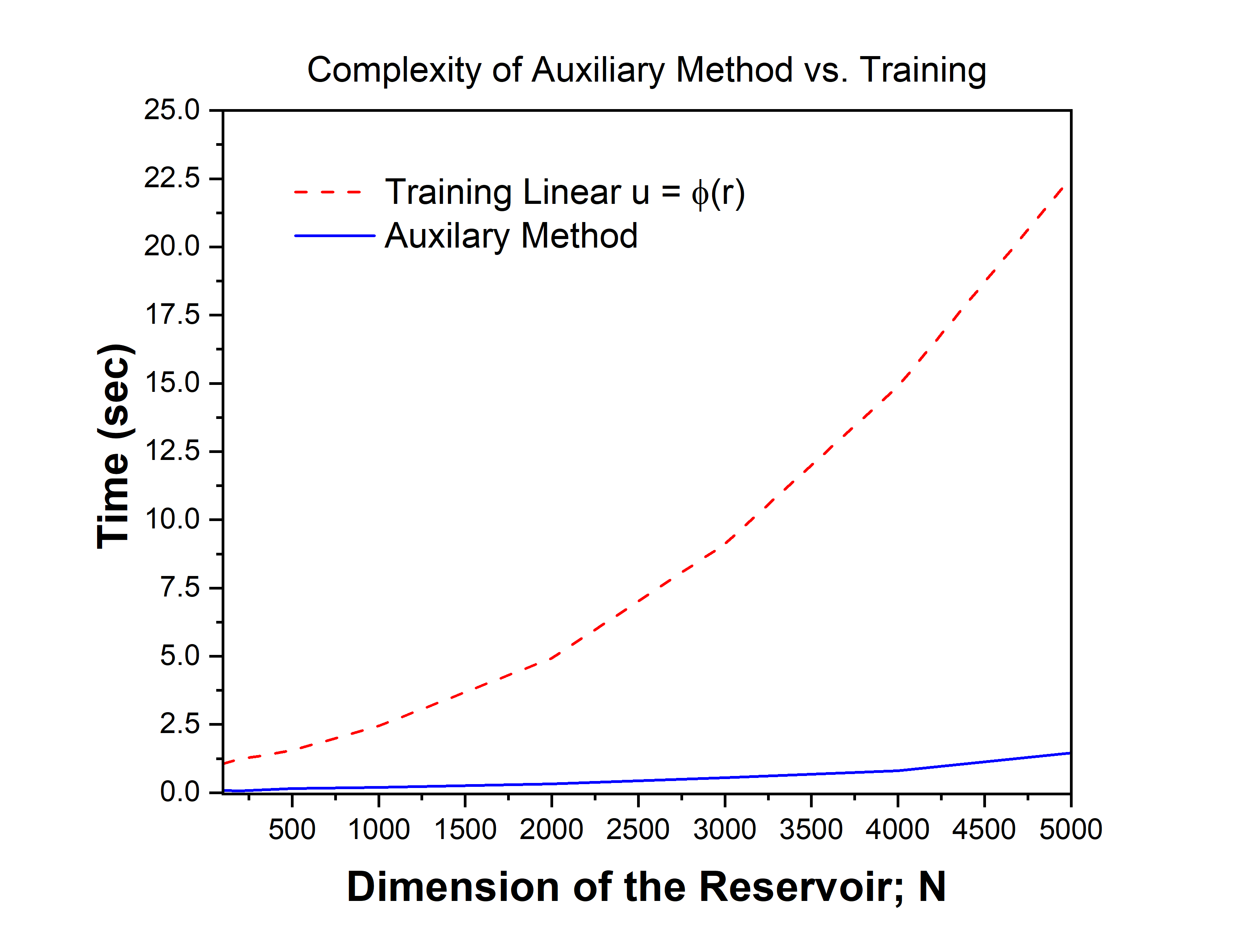}
    \caption{Time for testing a single set of hyperparameters for a tanh RC as a function of reservoir dimension using either the auxiliary method or directly through training the linear regression output.  The test was executed on a personal desktop with an intel i9-9900K cpu.  We recommend using the auxiliary test to find the bounds of the feasible space of hyperparameters for the RC.  In addition, if using a cost function based optimization, then one could use the auxiliary method as a prescreen for a proposed set of parameters.}
    \label{fig: complexity}
\end{figure}

As an example of the computational speedup afforded by the auxiliary method test, we evaluated the timing of the test vs the timing of training an RC as a function of reservoir dimension---Fig.(\ref{fig: complexity}).  The benefit in using the auxiliary test comes for large N RCs.  

Even greater benefit is accrued when running on a big system using a parallel reservoir scheme on localized patches such as in section 3.2.3.  Generally PGS on one of the parallel RCs implies PGS on all so we can decrease the computation substantially.  The test for the SWE in 3.2.3 took 16 seconds per point searched, while training required a several hundred node CPU cluster with multiple reservoirs running in parallel and each point in the search required 20 minutes.  In a Bayesian optimization setting such as in \cite{Griffith19} this test can be used as a binary (PGS or no PGS) prescreening function to the cost function evaluation.

As may be seen in Fig.(\ref{fig: AUXandCLE_FHN_pnzSR}), the boundary between PGS and nonPGS regions is, in practice, quite sharp. This also seen in Fig.(\ref{fig: robust}). So the search for PGS and nonPGS regions in $\btheta$ space may be coarse grained to begin and subsequently refined if desired~\cite{storn97}.  Additionally, it has been suggested that RC works best at the ``edge of stability'' \cite{Carroll20}.  The auxiliary method allows the user to find this ``edge'' and concentrate the search for the hyperparameters in the border region where it has empirically been shown that the RC works best.

Testing for PGS only tells us that the function $\bvarphi(\r)$ exists, not whether our approximation to it is sufficient for prediction.  One would expect a linear approximation in $\r$ to $\bvarphi(\r)$ would predict well only for a small subset of the parameters for which PGS is shown to occur; indeed this is exactly what we find empirically.  We suggest that richer approximations for  $\bvarphi(\r)$  might expand this subset of good predictions to include much more of the region indicated by the PGS test.

\section{Using PGS in Forecasting Examples}

%Fig.(\ref{xlor63tores}) shows the result of using RC for a reservoir with $\tanh$ activations units driven by data from the 3 dimensional equations of Lorenz~\cite{lor63}, with and without established PGS.

\subsection{Two RC Networks}
For the purposes of this paper, we have used two quite distinct dynamical systems at the reservoir nodes. One has $\tanh$ nonlinearities at the reservoir nodes, and the second has nonlinear FHN~\cite{fitzhugh1961impulses,Nagumo1962} oscillators at its nodes. Our methods are the same, giving qualitatively similar results, for the identification of PGS and noPGS regions for these and many other instantiations of input signals and reservoir activation functions. We have used simple Hodgkin-Huxley neuron models~\cite{hodgkin1952quantitative,jwu,willshaw} in an RC architecture, but do not report the results in this paper.

These simple and familiar choices emphasize our earlier comment that any smooth dynamics may be used as activation functions in an RC.

The descriptions of these two choices now follows:

\subsubsection{The Hyperbolic Tangent RC}

We use the adjacency matrix $\bf{A_{\alpha\,\beta}}$ and nonlinear activation function $\tanh()$ in this RC model. $\gamma$ adjusts the time scale of the reservoir dynamics, and $\sigma$ weights the strength of the input signal $\u(t)$
\be
\frac{d\r_\alpha (t)}{dt} = \gamma \biggl[-r_{\alpha}(t) +\tanh(A_{\alpha \beta}r_{\beta}(t) + \sigma W_{\alpha a}u_a(t))\biggr].
\label{tanhres}
\ee
Repeated indices are summed over. %This is not the only formulation that is possible for $\F_r$. In the Appendix we discuss other RC strategies for reservoirs including those based on nonlinear neuron models as active units. 
The $N \times D$ matrix $W_{\alpha a}$ determines where the input signals are introduced into the reservoir.

In Table (\ref{tab: hyperparams}) we display the meaning of the parameters in the $\tanh$ RC. For other selections of active units and RC architectures different tables may be useful.
 \vspace{0.2in}
\begin{table}[htpb!]
\centering
\begin{tabular}{|c c|} 
 \hline

 {\bf Parameter} & {\bf Description} \\ %[0.5ex] 
 \hline\hline
 SR & Largest eigenvalue of the adjacency matrix $A_{\alpha\,\beta}$ \\ 
  \hline
 $\rho_A$ & Density of the adjacency matrix $A_{\alpha\,\beta}$ \\
  \hline
 N & Degrees-of-Freedom of the reservoir\\
 \hline
 $\gamma$ & Time constant of the reservoir computer \\
 \hline
 $\sigma$ & Strength of input signal \\ %[1ex] 
 \hline
\end{tabular}
%\label{hyper}
\caption{{\bf Parameters and hyperparameters of the tanh reservoir computer, Eq. (\ref{tanhres})}. Other reservoirs have various hyperparameters that depend on the active units at their nodes, the adjacency matrix specifying their connection, and $\btheta$ in the reservoir vector field $\F_r(\r,\u,\btheta)$.} 
\label{tab: hyperparams}
\end{table}

\subsubsection{The FHN RC}

The equations for the Fitzhugh-Nagumo Model (FHN)~\cite{fitzhugh1961impulses,Nagumo1962} operating at reservoir sites $\gamma = 1,2,..,N$ are 

\be
\frac{dV_{\gamma}(t)}{dt} = \frac{1}{\tau}[V_{\gamma}(t) - \frac{1}{3}V_{\gamma}(t)^3 - w_{\gamma}(t)] 
%+\sum_{\nu \ne \gamma}^N I0(\gamma,\nu) 
+ \biggl\{\sum_{\beta = 1}^N A_{\gamma\beta}V_{\beta}(t)I0(\gamma,\beta) + \sum_{b=1}^D W_{\gamma b} u_b(t) \biggr\} 
\ee
\be
\frac{dw_{\gamma}(t)}{dt} = V_{\gamma}(t) - \eta w_{\gamma}(t) + \xi \nonumber
    \label{eq: FHN}
\ee 
%with $i$ denoting post-synaptic, $j$ pre-synaptic
The constants are  $\xi = 0.7$, $\eta = 0.8$, $\tau = 0.08$ ms. $I0(\gamma,\beta) = \frac{I_0}{2}[1+\tanh(K(V_{\beta}(t)-V_p))]$ is a synaptic current from a presynaptic neuron labeled by $\beta$ to a postsynaptic neuron labeled by $\gamma$; $K = 3/2$, $V_p = 1$, $I_0 = 1.0$.

\subsection{Data Sources}

\subsubsection{Lorenz63 Model}

The Lorenz-63~\cite{lor63} equations form a deterministic nonlinear dynamical system that exhibits chaos for broad ranges of parameters. It was originally presented as a three dimensional, {\bf very} reduced, approximation to the partial differential equations for the heating of the lower atmosphere of the earth by insolation. The dynamical equations of motion are
\bea
&& \frac{dx(t)}{dt} = \sigma[y(t) - x(t)] \nonumber \\
&&\frac{dy(t)}{dt}     = x(t)[\rho - z(t)] - y(t) \nonumber \\
&&\frac{dz(t)}{dt} = x(t)y(t) - \beta z(t)\\
\label{lor63}
\eea
with time independent parameters often chosen to be $\sigma = 10, \rho = 28, \beta = 8/3$.

The global Lyapunov exponents here are $\{\lambda_1, \lambda_2,\lambda_3\} = [0.9,   0,  -14.7]$ calculated using the QR decomposition algorithm given by Eckmann and Ruelle~\cite{eckmann85}.

We chose this model to provide three dimensional input $\u(t) = \{x(t),y(t),z(t)\}$ from the Lorenz63 system, as it is a test bed for many new ideas involving chaotic time series.

\subsubsection{Lorenz-96 Model}

This model of Lorenz~\cite{lor96,lorman98} is widely used in geophysics to examine new methods of data assimilation~\cite{abar2021}. It has the structure of `stations'' at $x_a(t)$ on a ring forced by a constant $f$.
For $f \approx 8.0$ the $\x(t)$ are chaotic when $D \ge 3$.

The dynamical equations introduced by Lorenz~\cite{lor96,lorman98}:
\be 
\frac{dx_a(t)}{dt} = x_{a-1}(t)(x_{a+1}(t) - x_{a-2}(t)) - x_a(t) + f
\label{Lorenz-96}
\ee
and $a=1,2,...,D$; $x_{-1}(t) = x_{D-1}(t)$; $x_0(t) = x_D(t)$; $x_{D+1}(t) = x_1(t)$. $f$ is a fixed parameter which we take to be in the range 8.0 to 8.2 where the solutions to these dynamical equations are chaotic~\cite{kostuk,abar2021}. The equations for the states $x_a(t);\; a = 1, 2, ..., D$ are meant to describe `stations' on a periodic spatial lattice. We use $D = 5$.

The Lyapunov exponents are $\{\lambda_1, \ldots,\lambda_5\} = [0.6,  0.0, -0.4, -1.4, -3.8]$ and were evaluated via the QR decomposition algorithm given by Eckmann and Ruelle~\cite{eckmann85}.

Data from this model are used in Section (\ref{assess}).

\subsubsection{Shallow Water Equations} 

These are equations for three fields the East-West and North-South velocities, and the fluid height $[u(x,y,t), v(x,y,t), h(x.y.t)]$ = $[$\V(x,y,t)$, h(x,y,t)]$ and describe the flow of a thin ($\approx$ 10km) layer of fluid on a two dimensional mid-latitude plane tangent to the earth ($\approx$ 6400km in radius)~\cite{sadourny75,pedlosky1986,vallis17}. They form a core contribution to much more complex models of the atmosphere/ocean system in weather and climate prediction models.

\be
\frac{\partial \V(x,y,t)}{\partial t} + \eta (\hat{z} \times (h(x,y,t) \V(x,y,t))) +\nabla [gh(x,y,t) + \frac{\V(x,y,t)^2}{2}] = 0 \nonumber
\ee
and
\be
\frac{\partial h(x,y,t)}{\partial t} + \nabla \cdot (h(x,y,t) \V(x,y,t)) = 0.
\ee
$\hat{z}$ is a unit vector normal to the (x,y) plane of the fluid flow. g is the strength of gravity and
\be
\eta = \frac{\partial_x v(x,y,t)  - \partial_y u(x,y,t)}{h(x,y,t)},
\ee
is the potential vorticity.

%fig 4
\begin{figure}[!htpb]
\centering
\includegraphics[width=0.65\textwidth]{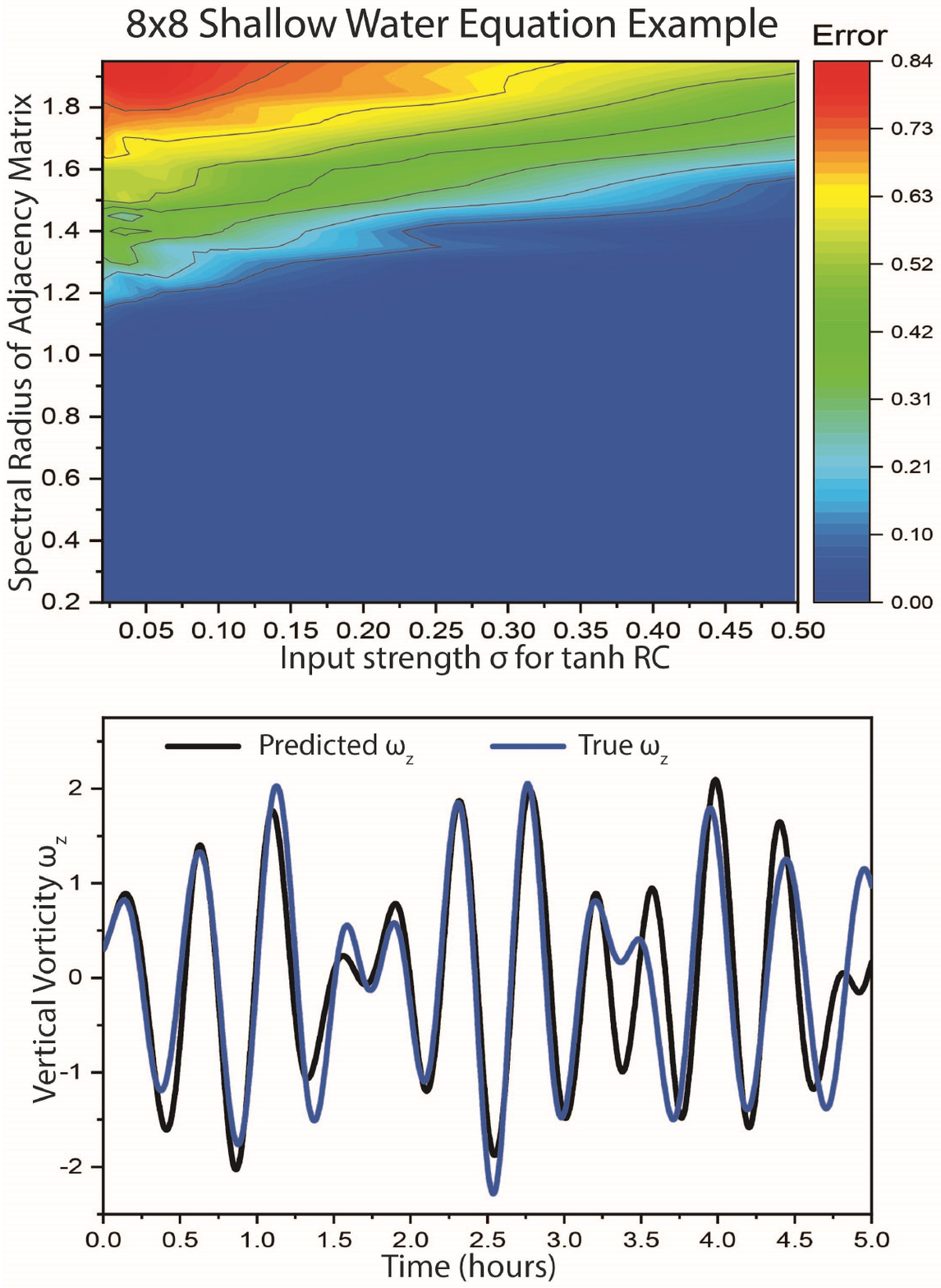}
\caption{{\bf Top Plot} Contour plot of {\bf PGS} and {\bf no PGS} regions. Blue/Purple indicates a region of parameters in a localized $\tanh$ reservoir model (N = 5000) which shows PGS or noPGS with a driving signal from the 8 $\times$ 8 Shallow Water Equations (SWE) \cite{sadourny75,pedlosky1986} as $\u(t) \in \mathbb{R}^{192}$. The red region shows no PGS.\\ {\bf Bottom Plot} Forecast for the normalized vertical vorticity at a particular point on the 192 dimensional $8\times8 \times 3$  grid.  The forecast starts here at time 0 after a spinup period which is not shown.  The localized reservoir algorithm and details of the SWE are found in section 3.2.3.}
\label{fig: swe}
\end{figure}

\subsubsection{Shallow Water Equation Results}
Accurate numerical solutions to the SWEs on a grid have been investigated in detail by Sadourny ~\cite{sadourny75} who concluded that a potential-enstrophy conserving scheme is effective. The details of this scheme can be found in Section 2 of ~\cite{sadourny75}. We use a form of the SWEs with three dynamical variables: surface height $\h(x,y,t)$, and the $\u(x,y,t)$ and $\v(x,y,t)$ components of velocity. We solve the SWEs numerically on a discretized grid of size $N_{\Delta} =8$ in two horizontal directions, resulting in an $8 \times 8$ grid. Including the three dynamical variables, this yields a D = 192-dimensional dynamical system. 

Following the scheme used in~\cite{pathak18} on a 1 dimensional grid, we use this discretized numerical integration of the SWEs to drive a set of localized reservoirs arranged in 16 overlapping local ``patches'' on a 2 dimensional grid. Each patch receives input from a subset of 48 local variables of the total 192-dimensional input vector.  The 48 variables input to each local reservoir consist of 16 $u(t)$, 16 $v(t)$ and 16 $h(t)$ that are located at the 16 points on a local patch of the grid. Each local reservoir is used to predict 12 (4 $u(t)$, 4 $v(t)$, 4 $h(t)$) of these after training, thus creating the overlapping scheme.

From the dynamical variables $\{u(x,y,t), v(x,y,t), h(x,y,t)\}$ we compare the reservoir output for normalized height and for the normalized vertical vorticity $\omega_z(x,y,t)$:
\begin{equation}
    \omega_z(x,y,t) = \frac{\partial v(x,y,t)}{\partial x} - \frac{\partial u(x,y,t)}{\partial y}.
\end{equation}
with their counterparts in the data. 

Even in this complicated set of overlapping localized RCs, it is straightforward and computationally efficient to apply our PGS test to the data.  Applying the auxiliary test we see---Fig.(\ref{fig: swe})---that there is a broad region where our 16 reservoir scheme synchronizes with the data.  This test is much more computationally efficient than evaluating the reservoir by training, thus giving us guidance as to where to focus our search.  Then, after a traditional search over this smaller grid of hyperparameters, a set of hyperparameters were found that produce reasonable and robust predictions over a short time scale.  The test enables us to significantly reduce the number of hyperparameters searched.

% Shown in Fig.(\ref{fig: swe}) is an example of determining PGS (or not) when a signal $\u(t)$ from the shallow water equations drive a tanh reservoir (N = 5000).  Here we wish to use RC to predict the evolution of the SWE on an $8 \times 8$ grid using a localized reservoir scheme, $\u(t) \in {\mathbb R}^{192}$.  Using the auxiliary method to test for PGS, we found a set of parameters for this high dimensional model quite efficiently.  
%See appendix for the detailed implementation details including the localized reservoir scheme.

We have also tested our procedures on other autonomous dynamical system drivers of an RC: (1) the Colpitts oscillator~\cite{creveling08} and (2) the Double Scroll system~\cite{gaut96}. The success of these investigations produces no additional guidance on using PGS, so we do not report on them any further in this paper.

\subsubsection{Driven Dynamical Systems; A Biophysical Example}\label{biophys}

The examples we have discussed until now have addressed data streams $\u(t)$ from autonomous dynamical systems as represented in Eq. (\ref{autondatasource}). Here we extend the discussion to a driven dynamical system.

A scientific area where RC may well find broad application is biophysics, in particular neurobiology. Neurons individually or in a functional biophysical circuit exhibit only a fixed point behavior (``resting potential'') when they do not receive external stimuli $\I(t)$. Clearly one does not need RC to predict the future of such a response.

When driven by electrical forces, usually a stimulating current $\I(t)$, the dynamics for the same neuron model~\cite{jwu,willshaw} is conditional on the forcing. The equations for the data stream change from Eq. (\ref{autondatasource}) to the non-autonomous data source system:
\be 
\frac{d\u(t)}{dt} = \F_u(\u(t), \I(t))),
\label{nonautondatasource}
\ee
and, as $\u(t)$ depends on the selected forces in $\I(t)$, information about this must be conveyed to the RC.
This physical setting for data sources is important well beyond biophysics.

As an example to show how RC works in this situation we have chosen to examine a data stream arising from the forcing of the simplest neuron models: that of Hodgkin-Huxley (HH). We use essentially the one they established 70 years ago~\cite{hodgkin1952quantitative,jwu,willshaw}.  The HH model is perhaps the earliest biophysical model of neuron oscillations. This neuron, a nonlinear electrical oscillator, has ion channels in its cell membrane through which Na$^+$ ions and K$^+$ ions flow, as well as a `leak' channel which represents the fact that the cell membrane leaks charge.

 We selected this model as it is the basis for the characterization of much more complex neuron structures.
 
 %fig 5
\begin{figure}[!htpb]
    \centering
     \includegraphics[width = 0.95\textwidth,height=0.5\textwidth]{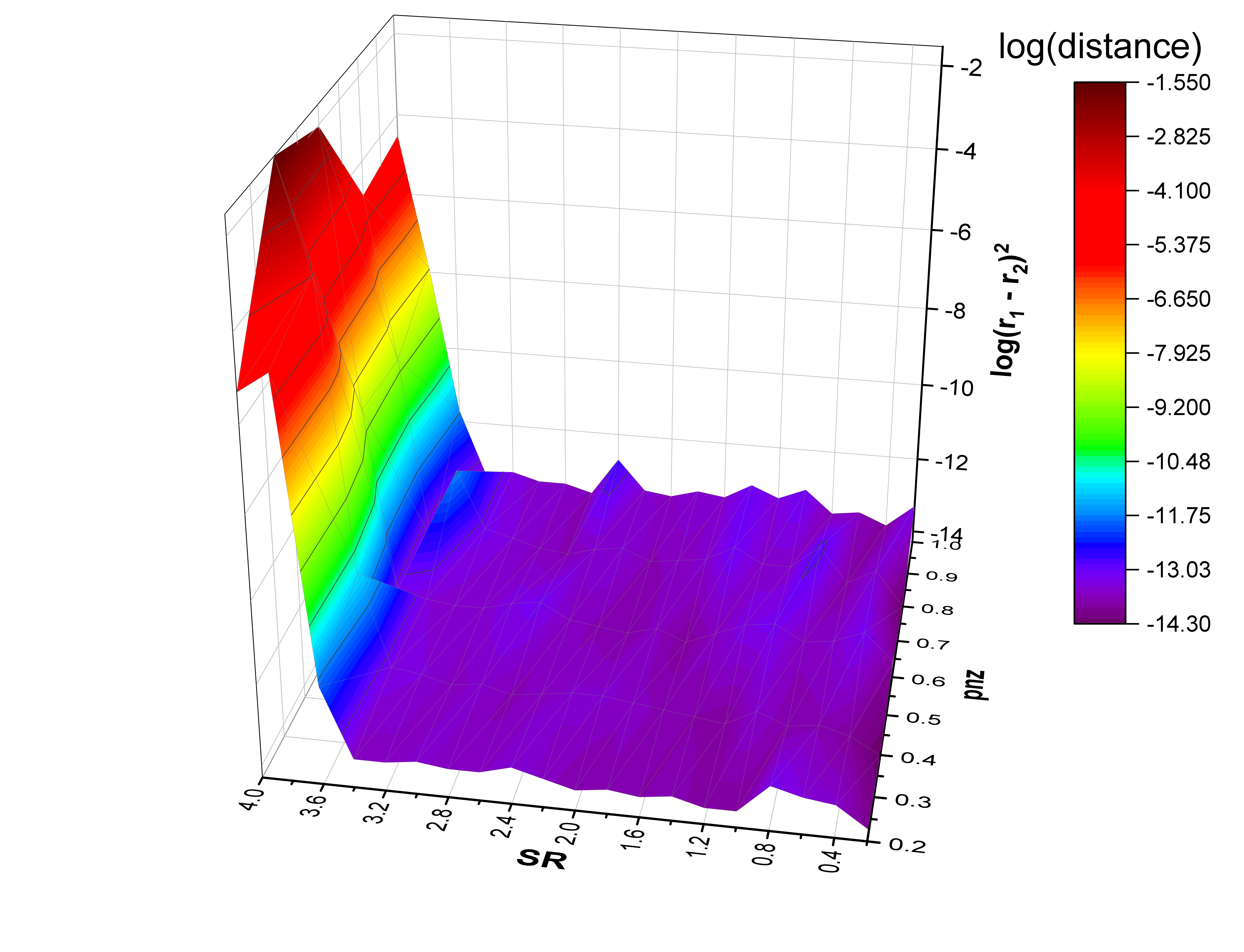}
    \includegraphics[width = 0.85\textwidth]{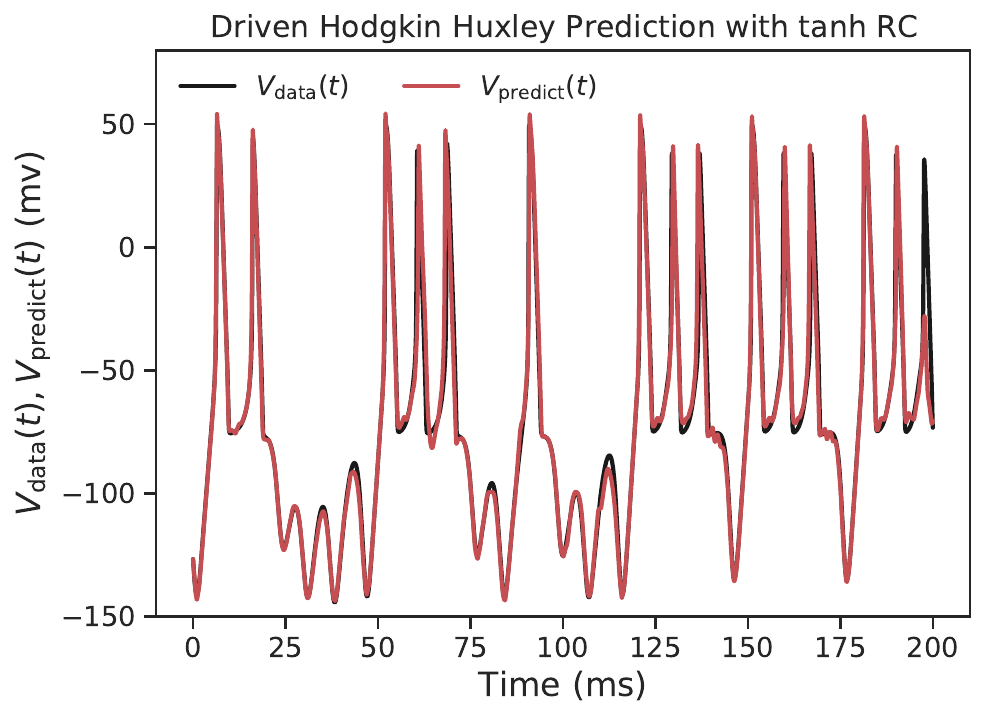}
    \caption{{\small {\bf Top Panel} Three dimensional display of the PGS and nonPGS regions in the case of an NaKL Hodgkin-Huxley neuron~\cite{hodgkin1952quantitative,jwu,willshaw}, driven by an external current with a waveform derived from the components associated with the Lorenz63~\cite{lor63} model, and presented to a $\tanh$ RC. These regions were selected using the auxiliary method, and we show the logarithm of $\|r_A(t) - r_B(t)\|$ for large times as a function of SR and $\rho_A$. The regions colored blue/purple are where PGS is found. The nonuniform surface reflects the residual roundoff error using single precision arithmetic in the calculations. {\bf Bottom Panel}  The cross membrane voltage showing both the known data and the prediction of the trained reservoir in the case of an NaKL Hodgkin-Huxley neuron~\cite{hodgkin1952quantitative,jwu,willshaw}, driven by an external current with a waveform derived from the components associated with the Lorenz63~\cite{lor63} model, and presented to a $\tanh$ RC. The parameters SR = 3.65 and $\rho_A = 0.02$ were selected to be in the PGS region of the top panel. }}
    \label{7dnakl}
\end{figure}
 
The Hodgkin-Huxley model is governed by the following four first-order differential equations
\bea
&&C\frac{dV(t)}{dt}=  I_{inj}(t) + g_{Na}m(t)^3h(t)(E_{Na}-V(t)) \nonumber \\
&& + g_{K}n(t)^4(E_K-V(t)) + g_L(E_L-V(t)). \nonumber \\
&&\mbox{\blank} \nonumber \\
&&\frac{dG(t)}{dt}= \frac{G_0(V(t)) - G(t)}{\tau_G(V(t))}\;\;\;\;\; G(t) = \{m(t), h(t), n(t)\}\nonumber \\
&&G_0(V)=\frac{1}{2}[1.0 +\tanh\left(\frac{V-V_G}{\Delta V_G}\right)]\nonumber \\
&&\tau_G(V)=\tau_{G0}+\tau_{G1}\left(1-\tanh^2\left(\frac{V-V_G}{\Delta V_G}\right)\right)
\label{HH}
\eea

In these equations the $\{g_{Na},g_K\}$ are maximum conductances for the Na and K ion channels, the $\{E_{Na}, E_K\}$ are reversal potentials for those ion channels, and 
$I_{inj}(t)$ is the external stimulating current injected into the neuron. The overall strength of an ion channel is set by the maximal conductances.

The gating variables $G(t) = \{m(t),h(t),n(t)\}$ are taken to satisfy first order kinetic equations and each lies between zero and unity, as they are effectively the probability that the ion channel is open. 

The quantities $G_0(V)$ and $\tau_G(V)$ are the voltage dependent activation function and the voltage dependent time constant of the gating variable $G(t)$. The forcing to the cell $I_{inj}(t)$, here a scalar, is known to us. The 19 parameters entering Eq. (\ref{HH}) are selected and further discussed in~\cite{toth2011dynamical} as well as in many other places~\cite{hodgkin1952quantitative,jwu,willshaw,abar2021}.

We, for this example, have chosen the forcing current to be a function $I_{inj}(t)$ taken to be proportional to the x(t) component of the Lorenz63 model. The Physics reasons for that are addressed in~\cite{toth2011dynamical}.

%Due to the statistical stationarity of the Lorenz63 model, we can predict both $I_{inj}(t)$ and the HH states forward in time.

We have found that, if we convey the data stream for the four Hodgkin-Huxley state variables, along with {\bf only} the x(t) component from the Lorenz63 model, then following our guiding path to working with an RC, we arrive at quite good predictions of the HH data stream (not shown here). However, if we add the information about the other Lorenz63 state variables, $\{y(t),z(t)\}$, the forecasting capability of the RC (here a $\tanh$ reservoir) is very much enhanced.

The procedure is to perform training and predicting with RC on the HH System (4 Dimensions) + L63 Dimensional System (3 Dimensions). The key difference between the driven and autonomous cases is that throughout the prediction process, the 3 dimensions of L63 are driven by their true values.  This means that the 4 dimensions of the NaKL system are linked together to the 3 dimensions of the L63 system through the mixing process in the reservoir.  The statistical stationarity of the L63 model allows predictions forward in time.

In Figure (\ref{7dnakl}) we show, in the {\bf top panel} the regions of PGS and nonPGS when a Hodgkin-Huxley data stream is presented to an N = 1000 dimensional tanh reservoir.  The parameters for the prediction in the \textbf{bottom panel} are chosen at the edge of stability (SR = 3.65) showing the benefit of first looking at the PGS regions and then searching for hyperparameters.

%In our parametrization of the cell dynamics there are $19$ fixed parameters and three unobserved state variables $a(t) = \{m(t), h(t), n(t)\}$ to be determined. All $a(t)$ lie between zero and one.

\section{Assessing Successful Reservoir Properties} \label{sec: eval}~\label{assess}
After running the test for PGS and performing a hyperparameter search, the question arises of how to guarantee stable forecasting.  One often encounters the two situations in RC:
\begin{itemize}
    \item The forecast starts out close to the data but then quickly diverges and becomes non-physical
    \item The forecast is ``good'' for certain initial starting conditions but not for others.
\end{itemize}

%fig 6
\begin{figure}[htpb!]
    \centering
    \includegraphics[width = 0.48\textwidth]{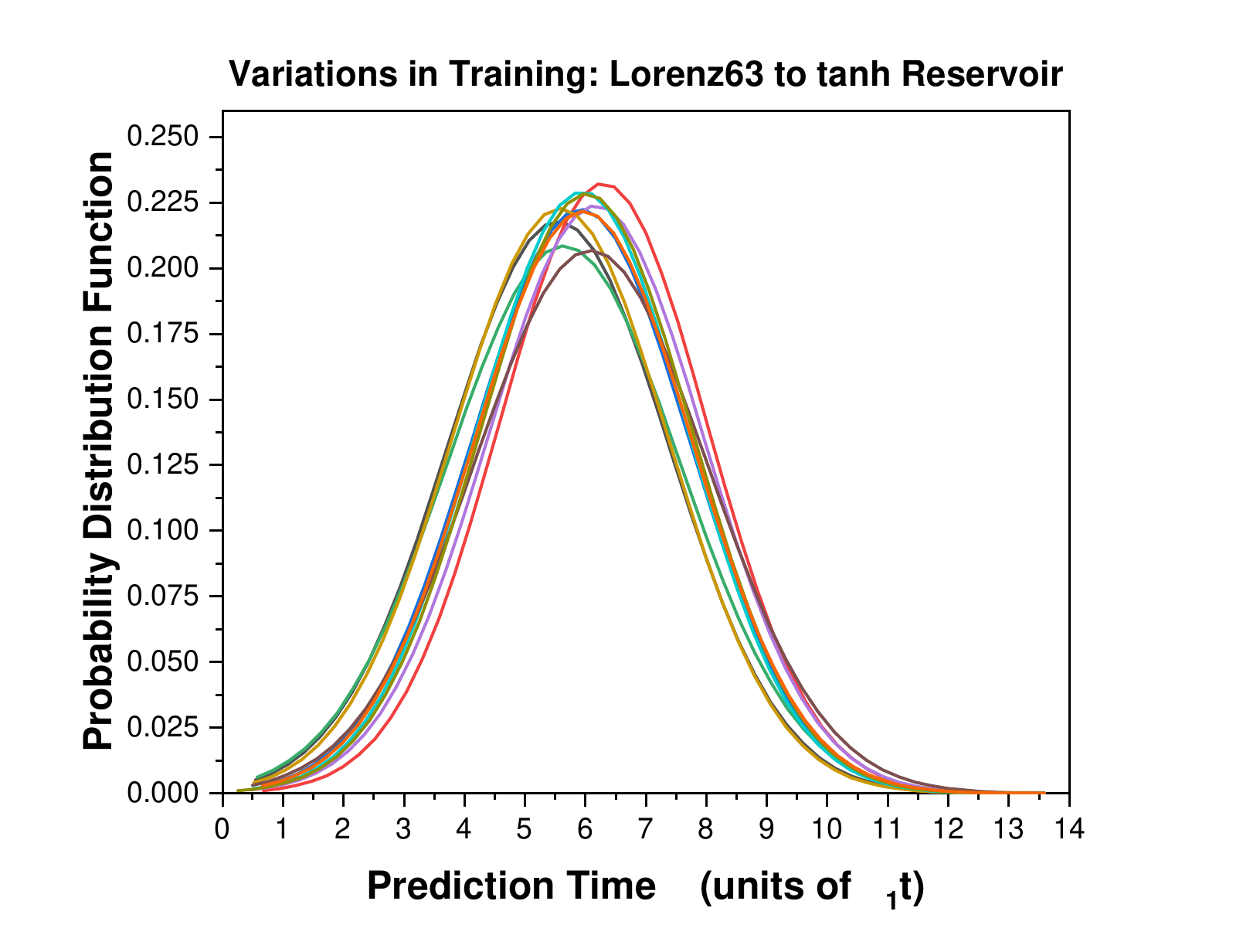}
    \includegraphics[width = 0.48\textwidth]{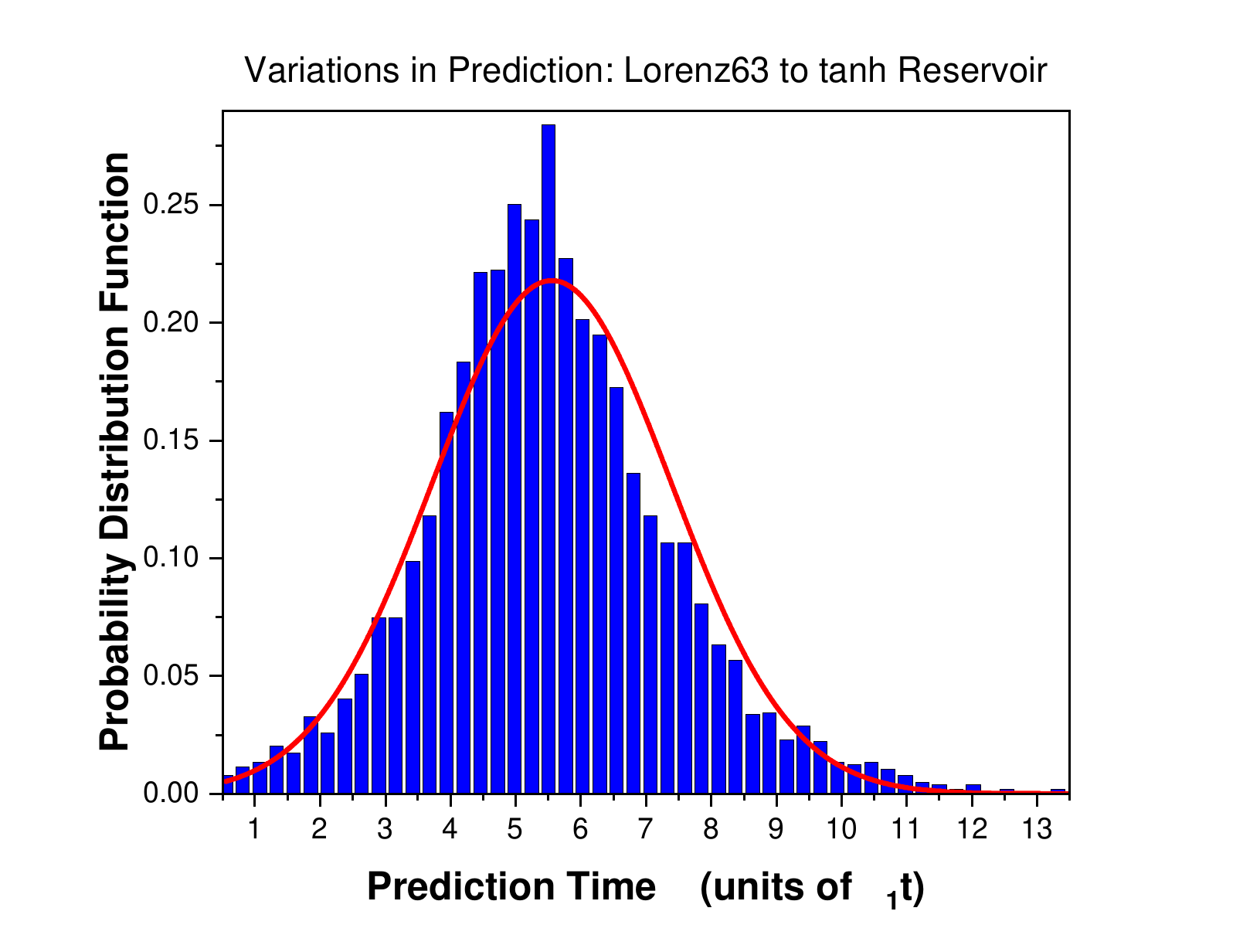}
    \caption{{\bf Left Panel} Gaussian fit to the prediction time for 10 N=2000 $\tanh$ RCs trained on Lorenz63 data with the same hyperparameters but different random seeds and training data.  The prediction time is the time for which the prediction stays close to the true values; the calculation is detailed in the appendix. Each reservoir predicts 4000 randomly selected training points. These points are different for each reservoir.  The 10 RC's prediction times overlap closely; the (mean(10 reservoirs)) = 5.92 and the (RMS deviation(10 reservoirs)) = 0.24.  This shows the robustness of this set of hyperparameters to training data and randomization of the reservoir layers. {\bf Right Panel} A histogram and the Gaussian fit to it from the {\bf Left Panel} better displays the variation shown in one hyperparameter setting of the reservoir computer.}
    \label{fig: robust}
\end{figure}

The problem of finding a set of hyperparameters that will give robust predictions is one of the main challenges in reservoir computing and RNNs in general.  The definition of a robust set of hyperparameters is one in which neither the randomness of the adjacency matrix $A_{\alpha\,\beta}$ or the training data set causes the reservoir to fail in prediction.  An example of robustness is shown in Fig.(\ref{fig: robust}) for a Lorenz63 input system.  Many sets of parameters work well for a particular example or point on the attractor. It is more challenging to find a set of parameters that predict well for many initial conditions and for different instantiations of the RC.
%\vspace{-0.2in}
%fig 7
\begin{figure}[htpb!]
    \centering
    \includegraphics[width = 0.65\textwidth]{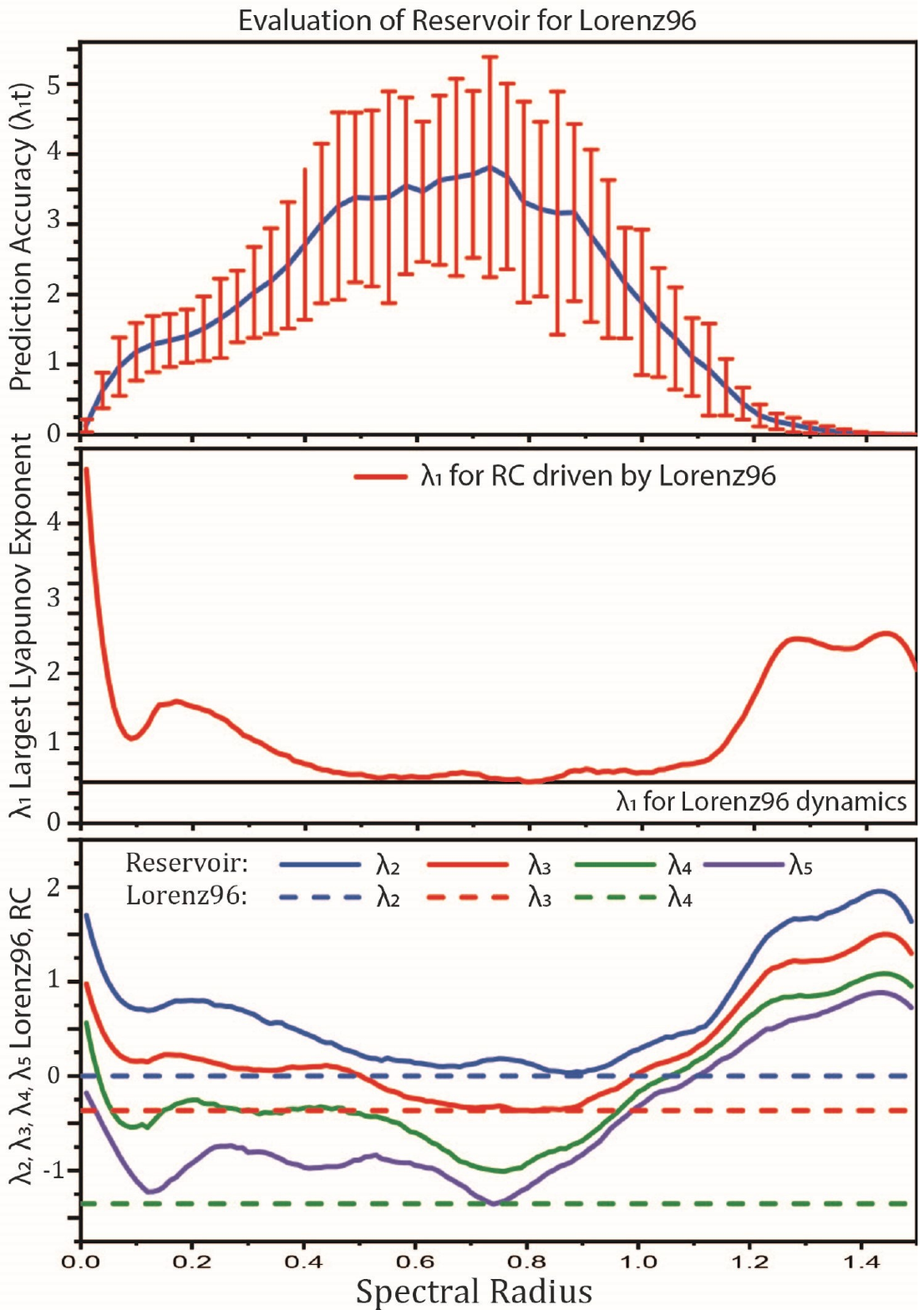}
    \caption{{\bf Top} Average prediction time of a N=2000 $\tanh$ reservoir as a function of SR when driven by a $D=5$ Lorenz96 $\u(t)$~\cite{lor96}. The time is in units of $\lambda_1 t$.  The error bars indicate variation in prediction depending on the stability of the input stimulus.
    {\bf Middle} The largest Lyapunov exponent, $\lambda_1$ of the forecast reservoir and $\lambda_1$ of the input system (black line) as a function of SR.
    {\bf Bottom} $\lambda_2,...,\lambda_5$. The next four Lyapunov exponents of the forecast reservoir . The method for computing the Lyapunov Exponents of an RC is discussed in  ~\cite{abar96,eckmann85,verstraten09}. Theoretically one of the LEs should always be 0, the slight discrepancy between 0 and the LEs is caused by numerical errors in estimating the LEs of a high dimensional dynamical system from a finite time sequence. Of course, we have displayed only a small subset of the LE's of the trained reservour computer in this Figure.}
    \label{fig: lyap_predlor96}
\end{figure}

%fig. 8
\begin{figure}[htpb!]
    \centering
    \includegraphics[width = 0.9\textwidth,height=0.65\textwidth]{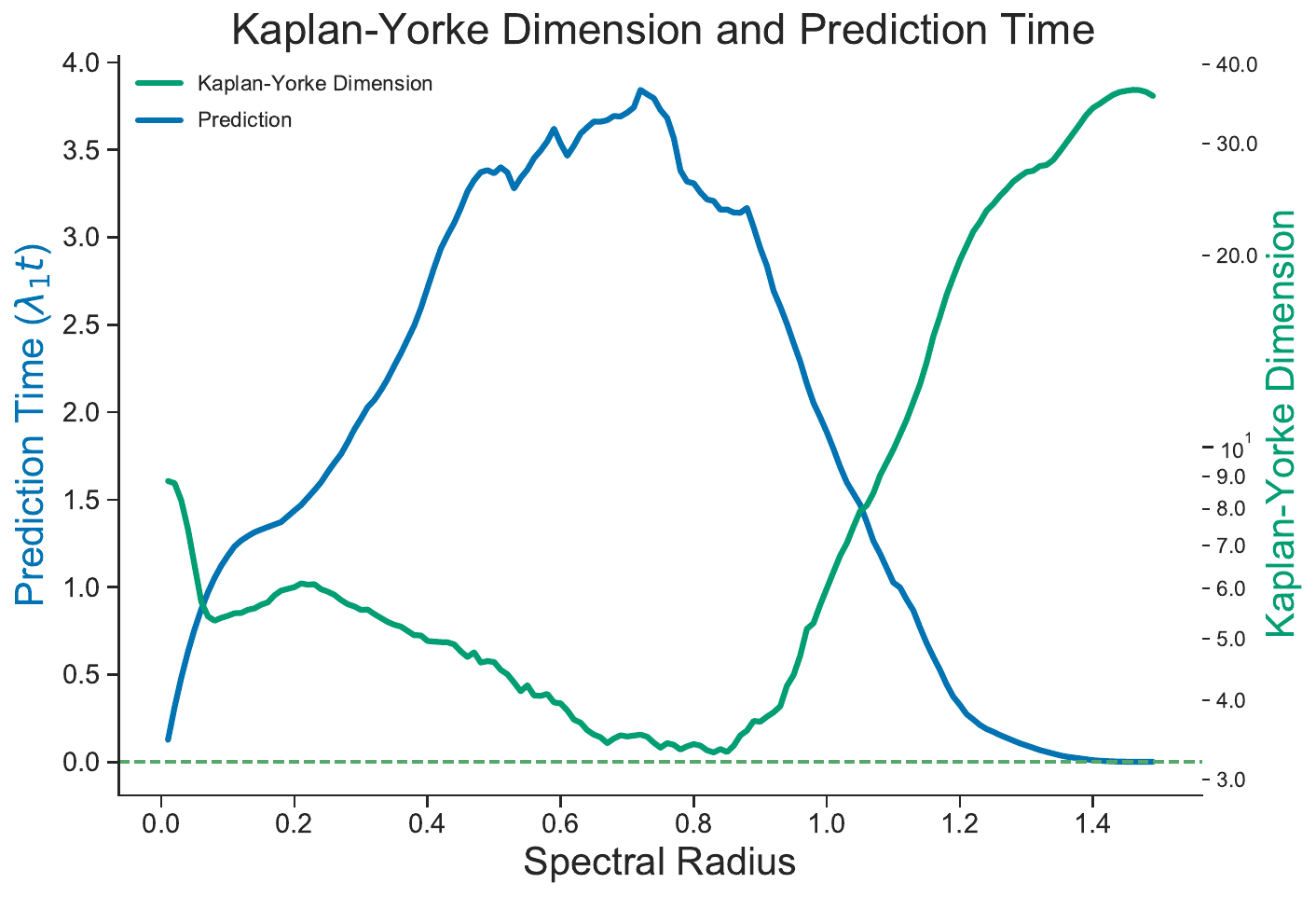}
    \caption{The Kaplan-Yorke fractal dimension~\cite{KaplanYorke79, KaplanYorke83} (Green) of the tanh forecasting reservoir (N=2000) trained by a Lorenz96 D = 5 system as a function of SR plotted along with the Prediction Time in units of $\lambda_1 t$ (Blue); this is the same system and RC as for the data shown in Fig.(\ref{fig: lyap_predlor96}).  The predicting reservoir KY dimension is an estimate of the dimensionality of the synchronization manifold where the RC resides.  The KY dimension of the 5D L96 system is shown as the dashed line.  As SR crosses $\approx 1.1$, corresponding to the largest Conditional Lyapunov Exponent of the reservoir crossing 0, the KY dimension of the reservoir increases rapidly.  This corresponds to the reservoir moving off the low dimensional PGS manifold. One expects that as this occurs, the forecasting capability of the RC will vanish quickly.}
    \label{fig: KYdim}
\end{figure}

The typical approach for evaluating ML predictions with the mean squared error over a test set does not capture a key feature of RC; a well trained RC should be able to give good short term predictions for {\bf all} initial starting points {\bf and} be stable in the medium to long term. This feature is called attractor reconstruction~\cite{hunt19}.  Instead of a test set, we propose an additional criterion for RC evaluation; {\bf a well trained RC reproduces the spectrum of Lyapunov exponents of the input system $F_u$}; Fig.(\ref{fig: lyap_predlor96}) for an example.

Lyapunov exponents (LEs) characterize the average global error growth rate of a dynamical system \cite{Lyapunov} along directions in phase space.  One can calculate the $N$ LEs of the forecast 
reservoir $\dot{\r}(t) = \F_r(\r(t), \bvarphi(\r(t),\btheta)$ and compare the largest of them to the $D$ exponents of the input system. If the $D$ largest LEs match and the smaller $N-D$ exponents of the RC are negative, then the two systems will have the same global behavior, increasing the likelihood of robust, stable predictions.  It is important to note that the exponents of the input system can be calculated directly from experimental data \cite{abar96}.

We show this calculation for the Lorenz96 system~\cite{lor96} in Fig.(\ref{fig: lyap_predlor96}). Our results show that when more of the spectrum of LE's are matched by the RC, the better the average predictions. In situations where it is difficult to exhaustively test the RC, perhaps because the model is expensive to run or there is limited data, evaluating the Lyapunov exponents of the forecasting reservoir will guarantee that the global error growth of the RC is the same as the data.  

%If the Lyapunov spectrum of the RC does not match that of the input then the two situations above are more likely to occur.  
%A similar calculation is performed in \cite{Lu18} but without systematically tying the results to the average prediction time.

The LE criterion should be used as a check on the robustness of the system; this is {\it in lieu} of checking robustness by training and predicting over thousands of initial conditions Fig.(\ref{fig: robust}). These data may not be available for experimental systems or simply computationally infeasible.

One could also use the matching of LEs as a cost function over which to evaluate the best hyperparameters. This calculation is, however, quite expensive on a cpu but may work more efficiently given a GPU implementation.  There are also reasonably efficient approximation techniques for estimating the largest exponent \cite{Rosenstein93}.

One can gain additional insight by plotting the forecast against the Kaplan-Yorke fractal dimension of the synchronization manifold Fig.(\ref{fig: KYdim}).  The KY dimension is calculated using the LEs of the autonomous forecast reservoir when it has moved onto the synchronization manifold. The lower the dimension of the synchronization manifold, the better the predictions will be. If there are too many degrees of freedom there is movement in directions that do not correspond to physically meaningful signals; thus the effect of the extra dimensions is to add the equivalent of noise into the reservoir. Another view is that a lower dimensional synchronization manifold corresponds to a smooth and continuous $\bvarphi$ which is quantified by the KY dimension.

We speculate that the lowest dimension of the synchronization manifold could be related to the minimum embedding dimension of the system \cite{abar96}---not necessarily the fractal dimension of those systems.  However, when calculated for the Lorenz 63 \cite{lor63}, Lorenz 96 \cite{lor96} and Colpitt's Oscillator we could find no direct correspondence between the dimension of the synchronization manifold and the embedding dimension.  Our failure in this regard does not mean that the correspondence does not exist.

The results presented in Fig.(\ref{fig: lyap_predlor96}) match the suggestion that the reservoir operates best at ``the edge of chaos''~\cite{verstraten09, Boedecker12, Jiang19}, that is, the maximal prediction time of the reservoir corresponds to a SR just less than 1.  \cite{Carroll20} makes the point that the ``edge of chaos'' is not necessarily always the best point for predictions of the reservoir.

\section{Conclusion and Discussion}\label{conclusions}

In this paper we have addressed the following topics:

\begin{itemize}
\item Generalized Synchronization for a dynamical system $\r(t)$ driven by a signal $\u(t)$
\be
\frac{d\r(t)}{dt} = \F_r(\r(t),\u(t),\btheta),
\ee
in which category Reservoir Computing (RC) belongs, has long been seen as the condition that there is some function $\bpsi(\u) = \r$ connecting the network $\F_r(\r(t),\u(t),\btheta)$ coordinates $\r$ to the unidirectional drive signal $\u$.
After a discussion of the possibility that this may translate to a similar connection $\u = \bvarphi(\r)$, which we called Predictive Generalized Synchronization (PGS) we concluded, based on our present knowledge, that this connection need not be global inverses of each other. We suggested, for example, in the case of multistability~\cite{gelig78,mc85,efimov12} of the dynamics of $\r$ and/or $\u$, such a global relation might be difficult to establish.

To bypass this subtle mathematical issue, we used PGS, $\u= \bvarphi(\r)$ in the discussion of RC, as this is all that is required to investigate how we might focus on the manner in which RC forecasts so well.
\item We introduced a computationally efficient numerical test, based on PGS, and using the `auxiliary method', to guide hyperparameter selections in RCs resulting in very good forecasting.
\item We delineated the ideas for the use of PGS with a few simple low illustrative models~\cite{lor63,lor96,lorman98} presented as $\u(t)$ to an RC for forecasting, then we turned to an important geophysical model~\cite{sadourny75,pedlosky1986,vallis17}, and finished with a discussion of a biophysical model of neuron dynamics~\cite{jwu,willshaw}. The last item produces data from a driven dynamical system (the neuron), and the data depend on an injection of current to stimulate the neuron into interesting oscillations. The RC must obtain information about the driving force as it is trained.
\item We explored a metric for a ``well trained'' RC network using the reproduction of the input system's Lyapunov exponent spectrum.
\end{itemize}

\subsection{Issues to Address}

The discussion in this paper directs attention to at least these issues:
\begin{itemize}
\item It appears interesting to complete the investigation of driven systems such as the neuron model we considered in Section (\ref{biophys}). What information is required,in detail, about the forces stimulating the driven system of interest. This is especially of interest when the dynamics, if any, of those forces may not be known.
\item With reference to the previous item, one should investigate how we can use RC to forecast in a non-stationary environment~\cite{hegger00,kennel97},~\cite{kantz04} Chapter 13, and references therein, see  also~\cite{sugi12}. This may be an important question for analysis of observed data.
\item We used a polynomial representation for $\bvarphi(\r)$, Section (\ref{poly}), following a path drawn in the literature~\cite{hunt19}. However, we have also successfully used radial basis functions~\cite{silverman86,p-rbfmir-87,buhmann09,casdagli89,billings13,broom88,Scott05,guill98} in this regard. This material is not presented in this paper. 
\item Once parameters for PGS regions for a given $\u(t)$ and a selected\\ $\F_r(\r(t),\u(t),\btheta)$ are found, one may still wish to seek choices in those regions yielding the `best' forecasting, Section (\ref{assess}). In this regard, not presented here, we have found the algorithm, Differential Evolution~\cite{storn97}, to be quite helpful.
\end{itemize}

Finally, we address the question of why all this could be very interesting for the use of ML in forecasting Physical, Geophysical, Biophysical, and other observed data. It might seem that requiring the use of a high dimensional `reservoir' $\F_r(\r(t),\u(t),\btheta)$ to forecast low dimensional observed dynamics is not promising. However, from the work of~\cite{canaday18} on implementing reservoirs in hardware, it appears that one can build special purpose computers to forecast the future of observed data and provide a fast, rather easily realized forecast machine. The simplicity of many choices of activation functions at the nodes of the RC lends itself to realizing this in practice.

 \section*{Acknowledgments} We had many productive discussions, especially on the use of generalized synchronization, with Brian Hunt and Ulrich Parlitz. Conversations with Dan Gauthier on the use of hardware instantiations of a reservoir have been very constructive.Three of us (SGP, J. Platt, and HDIA) acknowledge support from the Office of Naval Research (ONR) grants N00014-19-1-2522 and N00014-20-1-2580. SGP acknowledges further support from the National Oceanographic and Atmospheric Administration (NOAA) grants\\ NA18NWS4680048, NA19NES4320002, and NA20OAR4600277. We also thank Lawson Fuller for many fruitful discussions and for editing this paper.
 
 \section*{Data Availibility}
 The computer code and the data used in the graphics shown in this paper is open source and will be freely available from the authors on reasonable request.

%\bibliography{bibliography_v3}

\newpage
\appendix
\numberwithin{equation}{section}
\section{Appendix}

\subsection{Polynomial Expansion of $\bvarphi(\r)$}\label{poly}
GS assures us that the dynamical properties of the stimulus $\u(t)$ and the reservoir $\r(t)$ are now essentially the same. They share global Lyapunov exponents~\cite{ose68}, attractor dimensions, and other classifying nonlinear system quantities~\cite{sushchik95}.

The principal power of PGS in RC is that we may replace the initial non-autonomous reservoir dynamical system 
\be
    \frac{d\r_{\alpha}(t)}{dt} = F_{\alpha}[\r(t), \u(t)],
\ee
 with an autonomous system operating on the synchronization manifold \cite{Pecora97} %Eq.\eqref{eq: res forecast}
\be
    \frac{d\r_{\alpha}(t)}{dt} = F_{\alpha}[\r(t), \bvarphi(\r(t))].
    % \label{predres}
\ee
In practice, the function $\u=\bvarphi(\r)$ is approximated in some manner,  through training, and then this is substituted for $\u$ in the reservoir dynamics. 
%\newpage 
In previous work on this~\cite{hunt19,ottdresden19} the authors approximated $\varphi(\r)$ via a polynomial expansion in the components $\r_{\alpha};\; \alpha = 1,2,...,N$, and used a regression method to find the coefficients of the powers of $\r_{\alpha}$. 

This means we write $u_a(t) =\varphi_a(\r(t)) = \sum_{\alpha, \beta = 1}^N J_{a\alpha}\; r_{\alpha}(t) + Z_{a\alpha\beta}\; r_{\alpha}(t) r_{\beta}(t) + \ldots $, and we evaluate the coefficients $\{\J,\Z, \ldots\}$ by minimizing with respect to the constant matrices $J_{a\alpha}$ and $Z_{a\alpha \beta}$

\begin{equation}
\sum_t \biggl [u_a(t) - \biggl\{ \sum_{\alpha, \beta = 1}^N J_{a\alpha}\; r_{\alpha}(t) + Z_{a\alpha \beta}\; r_{\alpha}(t) r_{\beta}(t) + \ldots \biggr\}\biggr ]^2 + \mbox{regularization term}.
\label{taylor1}
\end{equation}
See the discussions in~\cite{Tikhonov43,Phillips62,miller70,tikhonov77,Press-Flannery-2007-NumRecipes}.

The dimension of $J_{a\,\alpha}$ is $D$ by $N$. The dimension of $Z_{a\,\alpha\,\beta}$ is D by $\frac{N(N+1)}{2}$ as it is symmetric in $\{\alpha,\beta\}$. If one simplifies to keeping only `diagonal' terms in $\{\alpha\,\beta\}$, then the second term in Eq. (\ref{taylor1}) 
is $\Z_{a\,\alpha} [r_{\alpha}]^2$ and this has dimension $D$ by $N$.

We use this polynomial representation for $\bvarphi(\r)$, noting there are many ways of approximating multivariate functions of $\r$~\cite{silverman86,buhmann09,casdagli89,billings13,broom88,Scott05}.

\subsection{What if only one component of the data is known ?}~\label{tdelay}

%First for autonomous system where we measure the scalar $s(t_n) = s(n)$

If we only know one component of the time series, say $s(t_n) = s(n)$, which is a scalar, we can define an M-dimensional proxy space of vectors $\S(n) \in {\mathbb R}^M$ via time delays~\cite{takens81,abar96,kantz04} as
\bea
&&\S(n) = [s(n), s(n-\tau), ..., s(n-(M-1)\tau)]  \nonumber \\
&& = [S_1(n),S_2(n),...,S_M(n)] 
\label{timedelay1}
\eea

From the definition of the components of $\S(n) = \{S_k(n) = s(n - (k-1)\tau)\}$ 
for $\S$ $\in$ ${\mathbb R}^{M}$ we have in time steps of $\tau$,
\be
\S(n+1) = \H(\S(n)),
\ee
We know something about $\H(\S)$: noting that $S_k(n+1) = S_{k-1}(n)$ for k = 2,...,M. For $k = 1$,  $S_1(n+1) = s(n+1) =  H_1(\S(n))$. $H_1(\S)$ is a scalar function of the M-dimensional variables $\S$.

We have, for $H_1(\S)$, the same problem we addressed in representing a function of many variables $\bvarphi(\r)$. This is an easier issue than representing $\bvarphi(\r)$, as it is a vector in 
$\mathbb{R}^{D}$, while  $H_1(\S)$ is a scalar.

To `train' $H_1(\S)$ we note that $S_1(n+1) = s(n+1)$, so $s(n+1) = H_1(\S(n))$. 
We know the values of s(n), so we can `train' $H_1(\S(n))$ by expanding it in a Taylor series in $\S(n)$, for example, and then determine the expansion coefficients in that series expansion as we did in Eq. (\ref{taylor1}) for $\bvarphi(\r)$. 

It is important to keep nonlinear terms in $\S(n)$ in the representation of $H_1(\S(n))$. %{\bf MORE HERE}
If we were to take $H_1(\S)$ to be linear, we would miss the nonlinear terms in the dynamical equations producing $s(t_n)$.
%~\cite{abar92}

\subsection{Prediction Quality} 
\label{sec: pred quality}
\begin{figure}[htpb!]
    \centering
    \includegraphics[width = 0.5\textwidth]{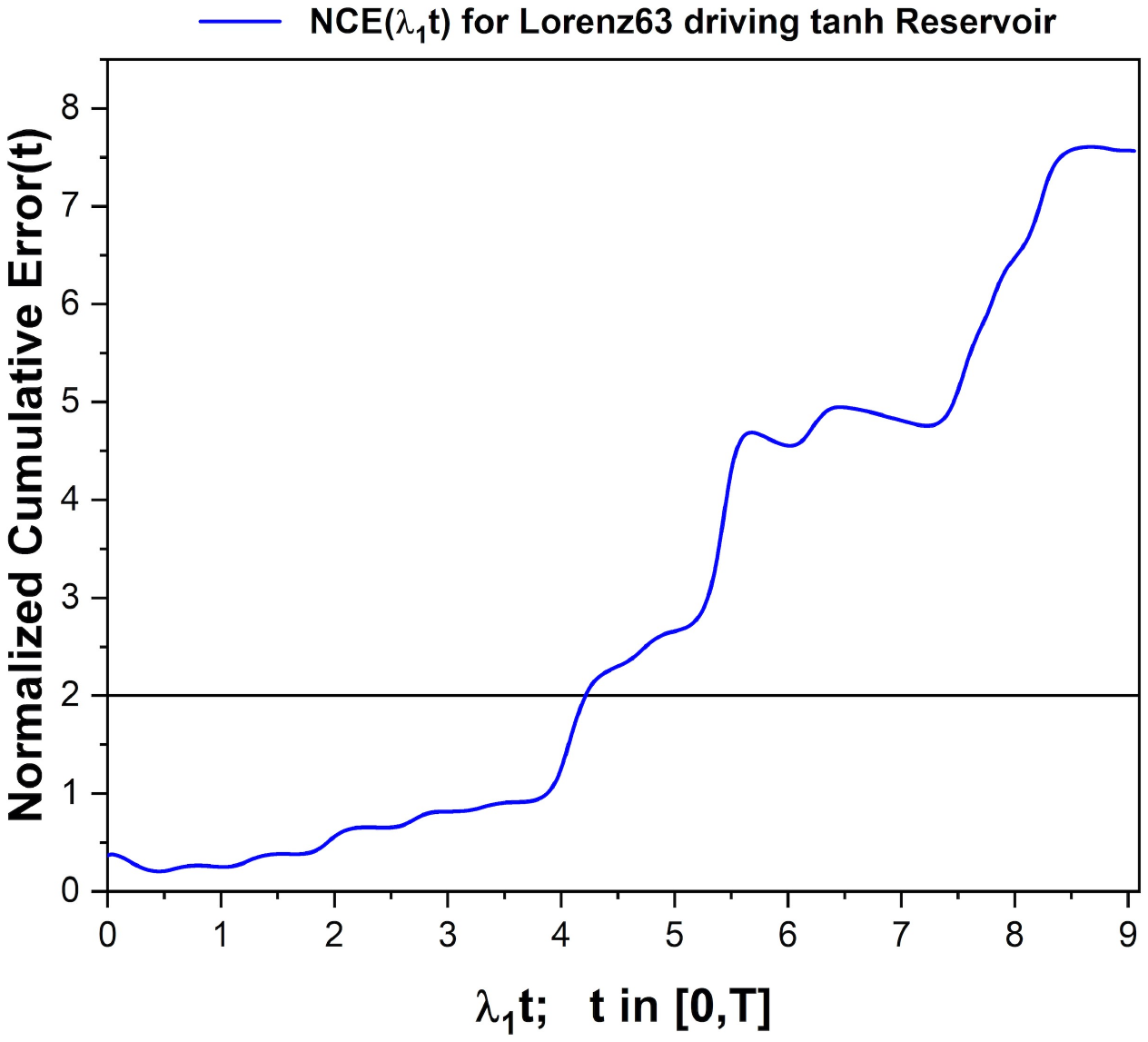}
    \caption{NCE($\lambda_1 t$), Eq.(\ref{norm cum error}), Lorenz63 input to a $\tanh$ RC; $N = 2000$.  NCE remains quite small for a long time. As the two time series separate NCE(t) rises. The dashed line suggests when the quality of the prediction has ``ended.''}
    \label{fig: error}
\end{figure}
To compare prediction times, a metric is required to estimate when the RC's capability to accurately predict $\u(t)$ ends. It needs to allow that the RC could be close to but not precisely the input for a number of time steps before diverging. A good metric for this {\bf Normalized Cumulative Error} (NCE):
\begin{equation}
    {\bf NCE(t)} = \sum_{t = 1}^T \frac{1}{t}[\u(t) - \varphi(\r(t))]^2.
     \label{norm cum error}
\end{equation}
$T$ is the length of the prediction phase. NCE($\lambda_1$t) stays small and flat for a long time before rising rapidly. 
One can see this in Fig.(\ref{fig: error}).

\newpage

%\bibliography{bibliography_v3}
%\bibliography{bookrefsnsf20.bib}
%\bibliographystyle{apalike}
%\bibliographystyle{alpha}
%\bibliographystyle{ieeetr}
\bibliography{chaos_resubmission1.bbl}

%\printindex
\end{document}